\documentclass[11 pt]{article}
\usepackage{amsfonts}
\usepackage{latexsym}
\usepackage[tbtags]{amsmath}
\usepackage{amsmath,amssymb,graphicx}
\usepackage{color}
\usepackage{graphicx}
\usepackage{graphics}
\usepackage{authblk}
\usepackage{cite}
\usepackage[usenames,dvipsnames]{xcolor}
\usepackage[pdfborder={0 0 0}]{hyperref}
\usepackage{algorithm} 
\usepackage{algorithmic}
\usepackage{lipsum}
\usepackage{tablefootnote}
\usepackage{fixltx2e}
\usepackage{enumitem}
\usepackage{multirow}
\usepackage{multicol}
\usepackage[flushleft]{threeparttable}
\usepackage{hyperref}
\hypersetup{
    colorlinks=true,
    linkcolor=blue,
    filecolor=magenta,      
    urlcolor=cyan,
}
\allowdisplaybreaks
\def\baselinestretch{1.1}

\def\inclde-picture #1 by #2 (#3){%
  \vbox to #2{%
  \hrule width #1 height 0pt depth 0pt
   \vfill
   \special{picture #3}}}
\def\scaledpicture #1 by #2 (#3 scaled #4){{%
  \dimen0=#1 \dimen1=#2
  \divide\dimen0 by 1000 \multiply \dimen0 by #4
  \divide\dimen1 by 1000 \multiply \dimen1 by #4
  \inclde-picture \dimen0 by \dimen1 (#3 scaled #4)}}

\def\inclde-picture #1 by #2 (#3){%
  \vbox to #2{%
  \hrule width #1 height 0pt depth 0pt
   \vfill
   \special{picture #3}}}
\def\scaledpicture #1 by #2 (#3 scaled #4){{%
  \dimen0=#1 \dimen1=#2
  \divide\dimen0 by 1000 \multiply \dimen0 by #4
  \divide\dimen1 by 1000 \multiply \dimen1 by #4
  \inclde-picture \dimen0 by \dimen1 (#3 scaled #4)}}
\newcounter{example}

\newcounter{para}
  
\newtheorem{definition}{Definition}[section]
\newtheorem{proposition}{Proposition}
\newcounter{remarkcnt}
\newenvironment{remark}
{
\stepcounter{remarkcnt} \noindent {\bf Remark \arabic{remarkcnt}.} }
\newcommand{\Lower}[1]{\smash{\lower 1.5ex \hbox{#1}}}
\newcommand{\HLower}[1]{\smash{\lower .03in \hbox{#1}}}
\newcommand{\LHigher}[1]{\smash{\raise .02in \hbox{#1}}}

\usepackage{natbib}
\usepackage{url}

\begin{document}

\setcounter{footnote}{1}

\vspace{.6in}


\title{\LARGE COVID-19: Optimal Allocation of Ventilator Supply under Uncertainty and Risk}


\author[1]{Xuecheng Yin}
\author[1]{\.I. Esra B\"uy\"uktahtak{\i}n\thanks{Corresponding author email: esratoy@njit.edu}}
\author[1]{Bhumi P. Patel}
\affil[1]{Department of Mechanical and Industrial Engineering, New Jersey Institute of Technology}

\maketitle

\renewcommand\baselinestretch{1.5}


\begin{abstract}
This study presents a new risk-averse multi-stage stochastic epidemics-ventilator-logistics compartmental model to address the resource allocation challenges of mitigating COVID-19. This epidemiological logistics model involves the uncertainty of untested asymptomatic infections and incorporates short-term human migration. Disease transmission is also forecasted through a new formulation of transmission rates that evolve over space and time with respect to various non-pharmaceutical interventions, such as wearing masks, social distancing, and lockdown. The proposed multi-stage stochastic model overviews different scenarios on the number of asymptomatic individuals while optimizing the distribution of resources, such as ventilators, to minimize the total expected number of newly infected and deceased people. The Conditional Value at Risk (CVaR) is also incorporated into the multi-stage mean-risk model to allow for a trade-off between the weighted expected loss due to the outbreak and the expected risks associated with experiencing disastrous pandemic scenarios. We apply our multi-stage mean-risk epidemics-ventilator-logistics model to the case of controlling the COVID-19 in highly-impacted counties of New York and New Jersey. We calibrate, validate, and test our model using actual infection, population, and migration data. The results indicate that short-term migration influences the transmission of the disease significantly. The optimal number of ventilators allocated to each region depends on various factors, including the number of initial infections, disease transmission rates, initial ICU capacity, the population of a geographical location, and the availability of ventilator supply. Our data-driven modeling framework can be adapted to study the disease transmission dynamics and logistics of other similar epidemics and pandemics.

\end{abstract}

\section{Introduction}
The world is undergoing a major health crisis, which has now eventually turned into a pandemic. The Coronavirus Disease 2019 (COVID-19), first detected in Wuhan city of China at the end of 2019, has been creating havoc on human life and economies in all parts of the world. Countries worldwide enforce lockdown and quarantine rules to slow down the spread of the virus. The lockdown, imposing travel restrictions, and social distancing have severely affected the economy, from small-scale industries to stock prices and international trading. The virus has such a high transmission rate, causing more than 104.7 million cases globally, out of which 2.3 million people have succumbed to death by mid-February 2021 \citep{JHU}. The continuous increase seen in coronavirus cases has made a worldwide scarcity of essential resources, such as ventilators, Intensive Care Unit (ICU) beds, Personal Protective Equipment (PPE), and masks. Effective, sufficient, and timely delivery of those critical resources to serve the COVID-19 patients has been a major challenge faced by the world countries during the pandemic.

COVID-19 is primarily an acute respiratory disease. Ventilator incubation delivers high oxygen concentrations while removing carbon dioxide and reduces the risk of hypoxia for COVID-19 patients \citep{meng2020intubation}. The standard Acute Respiratory Distress Syndrome protocol mandates that the most severe COVID-19 patients, who constitute $5\%$ of all COVID-19 patients, should receive ventilator support \citep{bein2016standard}. As a result, the life of many COVID-19 patients depends on the use of ventilators. The shortage of supplies and uncertainty in disease transmission has affected the proper allocation of ventilators, causing immense distress on the healthcare system. Due to ventilator shortages worldwide during the pandemic's peak times, hospital officials have had to make life-altering resource allocation decisions and prioritized the care of COVID-19 patients \citep{ranney2020critical}. To tackle ventilator shortages and reduce the number of COVID-related deaths, studies have come up with new approaches for ventilator distribution. For example, \cite{ranney2020critical} suggest that the demand for ventilators can be fulfilled by the government by allowing other industries to come together and help medical industries to cater to the needs of the ventilators. Another study by \cite{castro2020demand} suggests that the government in Brazil should start thinking about expanding the resource capacity rather than only focusing on the allocation of the available resources for controlling COVID-19. \cite{white2020framework} develop a framework for the distribution of ICU beds and ventilators depending on the priority scores using a scale of 1 to 8 based on patients' likelihood of survival and ethical considerations. 

Operations Research (OR) methods have been widely used to determine optimal resource allocation strategies to control an epidemic or pandemic. Several studies have used multi-period OR models to optimize the allocation and redistribution of ventilators (see, e.g., \cite{mehrotra2020model}, \cite{bertsimas2020predictions}, and \cite{blanco2020reallocating}). Other OR research models that study the epidemic diseases and resource allocation mainly focus on the logistics and operation management to control the disease in optimal ways \citep{buyuktahtakin2018new,zaric2001resource,
yinbuyuktahtakin2020,kaplan2003analyzing, tanner2008finding, cocsgun2018stochastic}. We refer the reader to excellent reviews of \cite{ dasaklis2012epidemics} and \cite{queiroz2020impacts} for a discussion of OR models for epidemic resource allocation.

While OR has been an extremely useful tool for effective and timely allocation of resources as a response to epidemics, none of the former work has considered the ventilator allocation problem using a risk-averse spatio-temporal stochastic programming model under uncertainty of asymptomatic infections. People move between regions, states, and countries, which aggravates the disease transmission to the other areas. Evaluating undetected or asymptomatic individuals is critical for determining disease dynamics because asymptomatic individuals move around and unknowingly infect other individuals  \citep{UCHealth}. Thus, the short-term migration of people is a critical factor that needs to be considered to forecast the transmission of the COVID-19 realistically. However, the short-term migration rate is hard to predict and is affected by interventions and human behaviors. Furthermore, disease transmission rates are not constant and rather evolve over time with government interventions, such as the lockdown or social distancing measures. This change in the transmission rates also should be considered in a realistic model. To our knowledge, none of the former OR ventilator allocation models have integrated the epidemiological aspects of the disease and resource allocation challenges in one optimization model.

In this paper, we address the limitation of realistically forecasting the transmission of COVID-19 and build a risk-averse multi-stage stochastic epidemics-ventilator-logistics programming model to study the ventilator allocation for the treatment of severe COVID-19 patients. Our model considers the uncertainty of untested asymptomatic individuals during the transmission of COVID19 and involves various pandemic scenarios for the proportion of untested infections during each time stage of the planning horizon. Our model also incorporates the short-term migration between the highly-impacted regions while using changing transmission rates under various non-pharmaceutical intervention measures. The model optimizes the distribution of ventilators while minimizing the total expected number of infected and deceased people. We calibrate, validate, and test our epidemiological ventilator allocation model using COVID-19 data collected during the pandemic's early stages.

\section{Literature Review}
This section presents a review of the articles that study mathematical models to estimate the transmission rate of COVID-19, evaluate various interventions used to control the disease, and optimize resource allocation during the pandemic. The literature covers various methods, including compartmental models, such as the Susceptible (S)-Exposed (E)-Infectious (I)-Recovered (R) [SEIR] formulations, simulations, optimization models, stochastic, statistical, or probabilistic approaches, and network or graph-theory models.

\textbf{Compartmental, Simulation and Network  Models.} Several studies on COVID-19 modeling use SEIR-type models with simulations, such as the Monte Carlo Simulation, to predict the disease transmission rate and analyze the effectiveness of proposed non-pharmaceutical interventions and resource allocation strategies (see, e.g., \cite{chatterjee2020healthcare}, \cite{weissman2020locally}, \cite{wang2020impact}, and \cite{li2020forecasting}). For example, \cite{li2020forecasting} develop a novel epidemiological model called DELPHI based on an SEIR model to estimate the effectiveness of government interventions and the effects of under-detection of confirmed cases. They find that the world-widely implemented travel restriction policies, social distancing, and mass gathering restrictions play a crucial role in reducing the infection rate. \cite{ku2020epidemiological} develop an empirical approach based on a Bass Susceptible-Infected-Recovered model to predict specific transmission parameters, exogenous forces of infection, and effective population sizes to determine the reproductive number ($R_0$) for the COVID-19 transmission in Chinese provinces.

Stochastic compartmental models have also been used to analyze the uncertain transmission dynamics of the COVID-19. For example, \cite{kucharski2020early} develop the stochastic dynamic Susceptible, Isolated, Infected, Recovered, and Exposed model to derive the transmission rate using the COVID-19 cases in Wuhan as well as the associated international infections. The initial reproductive number of 2.35 was reduced to 1.05 after imposing travel restrictions in Wuhan. \cite{kretzschmar2020impact} develop a stochastic mathematical model to understand the impact of time delays in testing and isolation on the reproductive number of the COVID-19 transmission. Differential equations have been mainly used to solve compartmental disease models (see, e.g., \cite{ambikapathy2020mathematical}, \cite{wei2020impacts}, \cite{zeb2020mathematical},    \cite{tuan2020mathematical}, and \cite{wang2020impact}). For example, \cite{roberts2020selectively} come up with a series of differential equations that are designed to compare the two scenarios based on delaying of ICU bed shortage - effects of hospitalizing fewer COVID-19 patients versus increasing the ICU bed capacity.

The literature also constitutes many studies that present statistical approaches, such as logistic regression, generalized additive, and time-series models to predict the growth of the disease over time and implement resource allocation strategies accordingly during the pandemic (see, e.g., \cite{murray2020forecasting}, \cite{manca2020simplified}, and \cite{katris2020time}). Graph-theoretical methods have also been quite useful in studying the COVID-19 transmission pattern and finding out the practices to reduce the transmission rate (see, e.g., \cite{hu2020risk}, \cite{loeffler2020covid}, and \cite{badr2020association}). For example,  \cite{badr2020association} model the transmission rate of COVID-19 based on mobile phone mobility. The study supports the role of social distancing in reducing the COVID-19 growth ratio, confirming that social distancing plays a major role in combating the COVID-19 unless vaccinations are made available worldwide.

\textbf{Optimization Models.} Optimization models have also been widely studied for resource allocation in the fight against COVID-19. \cite{ queiroz2020impacts} provide a systematic review of various supply chain and logistics approaches for optimizing the distribution of critical resources amid the COVID-19. To tackle the shortage of ventilators,  \cite{mehrotra2020model} develop a two-stage stochastic programming model, optimizing ventilator allocation during the pandemic under various demand scenarios. The authors find that when 60\% of the ventilator inventory is allocated to non-COVID-19 patients, there is no shortfall. In comparison, when 75\% of the stock is allocated to the non-COVID-19 patients, a shortfall in the supply of the ventilators to the COVID-19 patients occurs. Also, they find that it is essential to ramp up the production of the ventilators to meet the additional requirements of the ventilators that might come up during the peak times of the pandemic.  \cite{lacasa2020flexible} come up with an algorithm for optimizing the allocation of the ventilators and ICU beds and validate their algorithm during the peak and declining times of the pandemic based on the data from the United Kingdom and Spain cases. 

\cite{bertsimas2020predictions} develop a four-step approach, combining descriptive, predictive, and prescriptive analytics and propose an optimization model for the re-allocation of the ventilators throughout the U.S. during the COVID-19 pandemic. \cite{blanco2020reallocating} present a two-stage stochastic mixed-integer programming model, which minimizes the expected non-covered demand, using robust objective functions of type minmax and minmax regret. \cite{ billingham2020covid} present a network optimization model to tackle the problem of scarce ventilator distribution.  \cite{parker2020optimal} develop mixed-integer programming and robust optimization models to redistribute patients instead of resources, such as ventilators among different hospitals under demand uncertainty.  \cite{govindan2020decision} develop a practical decision support system hinge on the knowledge of the physicians and the fuzzy interference system (FIS) to help manage the demands of essential hospital services in a healthcare supply chain, to break down the pandemic propagation chain, and reduce the stress among the health care workers.

The literature on the optimal allocation of ventilators is not limited to COVID-19. For example, \cite{zaza2016conceptual} present a conceptual framework that identifies the steps in planning the distribution of stockpiled mechanical ventilators during an emergency. \cite{meltzer2015estimates} develop a spreadsheet model, which estimates mechanical ventilator demand in the United States during an influenza pandemic. They estimate a need of 35,000-60,500 additional ventilators to avert 178,000-308,000 deaths in a highly severe pandemic scenario. \cite{ huang2017stockpiling} introduce a two-stage method for optimizing stockpiles of mechanical ventilators, which are critical for treating hospitalized influenza patients in respiratory failure under a pandemic situation. They also incorporate their model into a web-based decision-support tool for pandemic preparedness and response.

\subsection{Key Contributions}
In summary, former stochastic programming approaches on ventilator allocation in a pandemic situation have involved a time domain of only two stages, and have not integrated an epidemic model within the stochastic program. Furthermore, the mathematical models on the forecast of the COVID-19 do not include the uncertainty of untested asymptomatic infections. They do not incorporate the impact of short-term migrations on COVID-19 transmission in an epidemiological model. Also, former studies on the COVID-19 modeling and logistics have omitted the time consistency of the risk for making decisions over multiple stages of a stochastic program under extreme pandemic scenarios. \\

Our modeling and applied contributions to the epidemiology and OR literature are summarized below.\\

\noindent \textbf{Modeling contributions:} First, to our knowledge, this is the first study that addresses the optimal distribution of ventilators to control a pandemic in a multi-stage stochastic mean-Conditional Value at Risk (CVaR) model. Considering multiple stages is essential to capture uncertain disease dynamic over multiple time periods.  This model includes many realistic effects critical in the COVID-19 pandemic, including untested asymptomatic infections, human movement among multiple regions, and evolving transmission rates under non-pharmaceutical intervention measures. Second, we consider the uncertainty of the proportion of untested asymptomatic infections at each stage and integrate this unknown dimension of the pandemic by generating a multi-stage scenario tree. Third, we present a new susceptible (S)- tested infected (I)- untested asymptomatic (X)- hospitalized (H)- ICU (C)- recovered (R)- death (D) compartmental disease model specialized for the COVID-19, and also incorporate the short-term human migration among multiple regions into this epidemiological model. Fourth, we derive a new time- and space-varying disease transmission formulation, which takes into account the impact of government interventions on transmission rates. Fifth, we formulate a budget-constrained ventilator allocation logistics model. Sixth, we incorporate a time-consistent CVaR risk-measure and the expectation criterion in the objective function to alleviate the impacts of extreme pandemic scenarios. Lastly, we integrate all those elements into one epidemics-ventilator-logistics mathematical formulation, which minimizes the number of infections and deceased individuals under different intervention strategies while determining the optimal timing and location of resources (ventilators) allocated. Our model combines the forecast of the transmission of COVID-19 and the determination of optimal ventilator allocation strategies in one formulation. Accordingly, the decision-maker can evaluate possible outcomes of wait-and-see decisions while foreseeing how the disease could progress in each time period.\\

\noindent \textbf{Applied Contributions:}
We apply our general multi-stage mean-risk epidemics-ventilator-logistics model to the case of controlling the COVID-19 in highly-impacted counties of New York and New Jersey. We collect real data from various resources and provide researchers with compact epidemiological, population and logistics-capacity data for COVID-19. Using this data, we calibrate, validate, and test our model, which could be used as a decision support tool for fighting against the COVID-19. Our model can also be adapted to study other similar diseases’ transmission dynamics and logistics.\\

\noindent \textbf{Key Recommendations to Decision Makers.} This study provides optimal risk-averse ventilator allocation policies under different risk levels that the decision-maker can take to control the COVID-19. Based on our results, we offer the following recommendations to inform resource allocation policies under a pandemic:

\begin{enumerate}[label={(\textit{\roman*})}]

\item The short-term movement of people influences the number of new infections even if the disease transmission rate stays the same.

\item The number of treated people in the ICU may stay at the capacity limit under different intervention strategies because this value depends on the minimum number of patients who require a ventilator for treatment and the scarce ventilator supply. There is also a lag time to observe the impacts of government non-pharmaceutical interventions on the number of hospitalized, ICU and deceased individuals.

\item ``Lockdown'' is the best strategy to control the COVID-19. However, the ``Mask and Social Distance'' intervention following a certain period of ``Lockdown'' is the second-best choice, considering the need for opening facilities and businesses.

\item The region with a high initial transmission rate and low initial ICU capacity will have more ventilators allocated under a limited budget and low or high transmission scenarios. Independent from the budget level, the region with a low initial transmission rate and low initial ICU capacity gets more ventilators allocated under a medium transmission scenario.

\item Under a medium and ample budget level, the model allocates more capacity to the regions with a higher population and a larger initial number of infections but with a lower transmission rate. A large-enough budget also provides some flexibility in delaying ventilator allocation to some regions. In contrast, all of the ventilators are allocated at the first two stages under a limited budget.

\item Considering risk in decision-making improves the confidence level for reducing the loss of lives under risky pandemic scenarios. However, a risk-averse decision-maker should also expect a possible increase in the number of infections and deaths while mitigating disastrous outbreak scenarios.

\end{enumerate}

\section{Problem Formulation}\label{Model}

This section presents the description of the notations, compartmental disease model, the formulation for transmission rates, uncertainty and scenario tree generation scheme, specific features and assumptions made in the mathematical model, a brief description of the CVaR, and the formulation for our epidemics-ventilator-logistics model.

\subsection{Model Notation and Formulation}

Below we provide notations used for the rest of the paper.

\begin{multicols}{2}

\textbf{Sets and Indices:}
\begin{enumerate}
\item[]$J$: Set of time periods, $J=\lbrace0,...,\overline{J}\rbrace$. 
\item[]$R$: Set of regions, $R=\lbrace1,...,\overline{R}\rbrace$.
\item[]$\Omega$: Set of scenarios, $\Omega=\lbrace1,...,\overline{\Omega}\rbrace$.
\item[]$N$: Set of nodes in the scenario tree, where $n \in N$.
\item[]$j$: Index for time period, where $j \in J$.
\item[]$r$: Index for region where $r\in R$.
\item[]$\omega$: Index for scenario, where $\omega \in \Omega$.\\

\textbf{State Variables:}

\item[]$S_{j,r}^{\omega}$: Susceptible individuals in region $r$ at stage $j$ under scenario $\omega$. 
\item[]$I_{j,r}^{\omega}$: Tested symptomatic infected individuals in region $r$ at stage $j$ under scenario $\omega$.
\item[]$X_{j,r}^{\omega}$: Untested asymptomatic infected individuals in region $r$ at stage $j$ under scenario $\omega$.
\item[]$H_{j,r}^{\omega}$: Hospitalized individuals in region $r$ at stage $j$ under scenario $\omega$.
\item[]$C_{j,r}^{\omega}$: Individuals treated in the intensive care unit (ICU) in region $r$ at stage $j$ under scenario $\omega$.
\item[]$R_{j,r}^{\omega}$: Recovered individuals in region $r$ at stage $j$ under scenario $\omega$.
\item[]$F_{j,r}^{\omega}$: Deceased individuals in region $r$ at stage $j$ under scenario $\omega$.
\item[]$O_{j,r}^{\omega}$: Number of tested symptomatic infected individuals admitted to the hospital in region $r$ at stage $j$ under scenario $\omega$.
\item[]$\overline{I}_{j,r}^\omega$: Number of tested symptomatic infected individuals who cannot be admitted to the hospital due to limited capacity in region $r$ at stage $j$ under scenario $\omega$.
\item[]$\overline{C}_{j,r}^{\omega}$: Number of individuals admitted to ICU in region $r$ at stage $j$ under scenario $\omega$.
\item[]$K_{j,r}^\omega$: Number of hospitalized individuals not admitted to the ICU due to the limited availability of ventilators in region $r$ at stage $j$ under scenario $\omega$.
\item[]$U_{j,r}^\omega$: Number of cumulative ventilators (ICU capacity) in region $r$ at stage $j$ under scenario $\omega$.
\item[]$\breve{I}_{j,r}^\omega$: Number of infections caused by short-term migration in region $r$ at stage $j$ under scenario $\omega$. \\

\textbf{Parameters:}

\item[]$\lambda_{1}$: Recovery rate of tested symptomatic infected individuals in region $r$.
\item[]$\lambda_{2}$: The death rate of tested symptomatic infected individuals in region $r$.
\item[]$\lambda_{3}$: Hospitalization requirement rate of tested symptomatic infected individuals in region $r$.
\item[]$\lambda_{4}$: Recovery rate of the hospitalized individuals in region $r$.
\item[]$\lambda_{5}$: Death rate of hospitalized individuals in region $r$.
\item[]$\lambda_{6}$: Ventilator requirement rate of hospitalized individuals in region $r$.
\item[]$\lambda_{7}$: Recovery rate of ICU patients in region $r$. 
\item[]$\lambda_{8}$: Death rate of ICU patients in region $r$. 
\item[]$\lambda_{9}$: Recovery rate of untested asymptomatic individuals in region $r$. 
\item[]$\sigma_{1,j,r}$: Transmission rate of tested symptomatic infected individuals in region $r$ at stage $j$.
\item[]$\sigma_{2,j,r}^{\omega}$: Proportion of untested asymptomatic infections in region $r$ at stage $j$ under scenario $\omega$.
\item[]$T_{j,r}^\omega$: Hospital capacity in region $r$ at stage $j$ under scenario $\omega$.
\item[]$U_{0,r}$: Initial number of ventilators (ICU capacity) in region $r$.
\item[]$e_{1}$: Cost of each ventilator.
\item[]$\Delta $: Total budget for ventilators.

\textbf{Risk parameters:}

\item[]$\alpha$: Confidence level of value-at-risk, where $\alpha \in \left[0,1\right)$.
\item[]$\lambda$: Non-negative risk preference parameter or mean-risk trade-off coefficient.\\

\textbf{Risk variables:}

\item[]$\eta_{j}^{\omega}$: Value at risk for each stage $j$ under scenario $\omega$.
\item[]$z_{j}^{\omega}$: Value exceeding the value-at-risk at the confidence level $\alpha$ at stage $j$ under scenario $\omega$.\\

\textbf{Non-anticipativity parameters:}

\item[]$n$: The serial number of nodes in the scenario tree, where $n \in N$.
\item[]$t(n)$: The corresponding stage that node $n$ marked in the scenario tree.
\item[]$\beta(n)$: The set of scenarios that pass through node $n$.\\

\textbf{Decision variables:}

\item[]$y_{j,r}^{\omega}$: Number of ventilators allocated to region $r$ at the end of stage $j$ under scenario $\omega$.\\

\end{enumerate}
\vspace{\fill}
\end{multicols}

\subsection{Compartmental Disease Model Description}

\begin{figure}[H]
\centering
\includegraphics[width=6in]{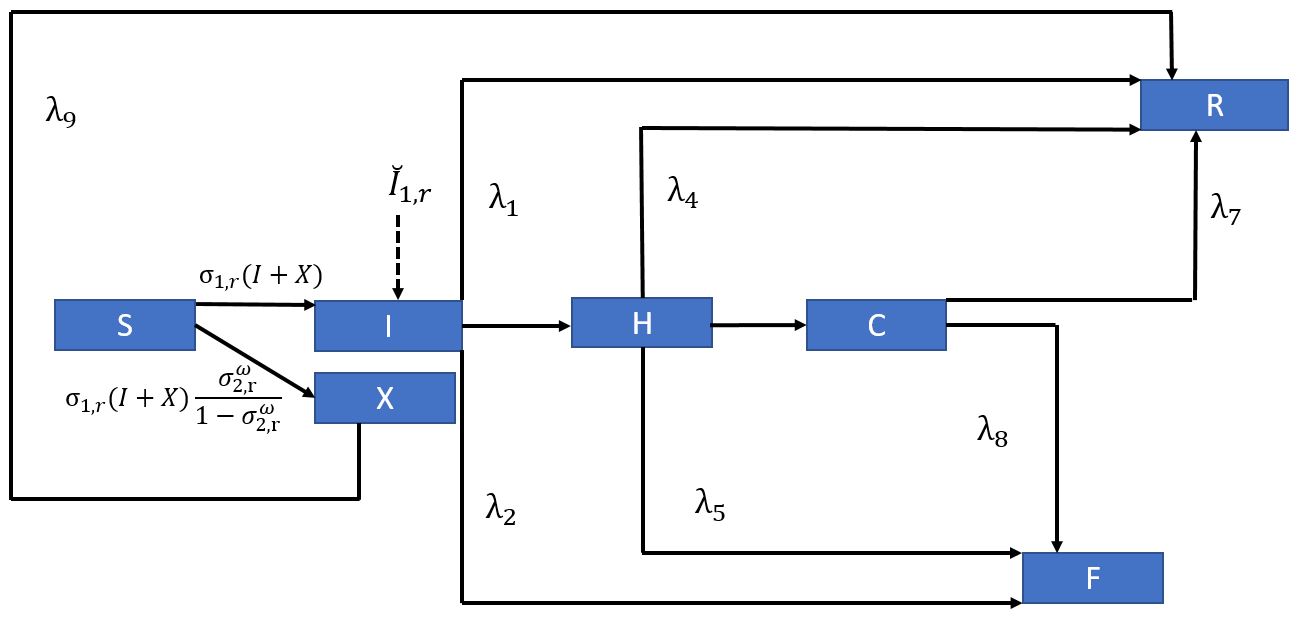}
\caption{One-Step COVID-19 Compartmental Model}
\label{Fig1}
\end{figure}

Figure \ref{Fig1} shows the transmission dynamics of COVID-19 in each region $r$ at each time period $j$ for a particular scenario $\omega$. In this figure, susceptible individuals (S) can be infected and become infected (either symptomatic or asymptomatic). Asymptomatic infections (X) may have slight or no symptoms throughout the infection period and will recover with a rate of $\lambda_9$. Tested symptomatic infections (I) may recover or die with rates of $\lambda_1$ and $\lambda_2$, respectively, if they are not treated in the hospital (H). Note that $\breve{I}_{j,r}^\omega$ with an incoming dashed arc to the I compartment represents the number of infected people coming into the region $r$ at stage $j$ under scenario $\omega$ from neighboring regions. Tested infected individuals (I) move to the hospital (H) compartment, depending on the number of tested infections (I) and available hospital capacity. Some of the treated infected people in the hospital (H) will recover with a rate of $\lambda_4$. The situations of some patients in the hospital (H) may get worsen, and thus they may be transferred into the intensive care unit (we use C to represent ICU), and those individuals need ventilators for the treatment.

Similar to the case of admittance into the hospital, the number of hospitalized patients transferred into ICU at each time period is equal to the minimum of the number of patients who need to be transferred into ICU and the number of available ventilators. The patients who are not able to receive the treatment in the ICU due to the limitation on the number of available ventilators may die at a rate of $\lambda_5$. After being treated in the ICU, some of the patients may recover with a rate of $\lambda_7$, while others may die with a rate of $\lambda_8$. Different from a typical compartmental model, the transfer rate from I to H and H to C is not a constant, and it depends on the available capacity in the H and C compartments, respectively, as discussed above.

\subsection{Time- and Space-Varying Transmission Rate}\label{TimeRate}

In this section, we formulate the transmission rate $\sigma_{1,j,r}$ as a time- and space-varying parameter, which depends on the government interventions taken at time $j$ and region $r$. Since the onset of the COVID-19, many governments have imposed different intervention strategies to reduce the transmission rate. At a certain stage, each intervention has a different impact on the transmission rate for the next stage. In this paper, we incorporate three main non-pharmaceutical interventions to formulate the time-varying transmission rate --- none, mask and social distancing, and lockdown.

Let $x_{j,r}^1$, $x_{j,r}^2$, and $x_{j,r}^3$ be binary decision variables that correspond to none ($i=1$), mask and social distancing ($i=2$), and lockdown ($i=3$) interventions, respectively, taken at stage $j$ and region $r$. If $x_{j,r}^i$ takes a value 1, then intervention $i$ is employed; otherwise, it is not employed, at stage $j$ and region $r$. The transmission rate at stage $j + 1$ in region $r$ is a function of the transmission rate and specific intervention employed at stage $j$ in the same region, as given in the below equations:\\
\begin{align}
\sigma_{1,j+1,r}=\sigma_{1,j,r}(m_{j,t}^1x_{j,r}^1+m_{j,t}^2x_{j,r}^2+m_{j,t}^3x_{j,r}^3) \quad \forall j\in J\setminus \lbrace \overline{J} \rbrace, \ r \in R,\label{timerate}\\
x_{j,r}^1+x_{j,r}^2+x_{j,r}^3=1 \quad \forall j \in J, \ r \in R,\label{decision}\\
x_{j,r}^1,x_{j,r}^2,x_{j,r}^3 \in \lbrace0,1\rbrace \quad \forall j \in J, \ r \in R,\label{binary}
\end{align}
where $m_{j,t}^i$ represents the percent change in the transmission rate with respect to the binary decision variable $x_{j,r}^i$ for intervention $i=1,2,3$ taken at stage $j$ in region $r$. Equation \eqref{timerate} shows that the transmission rate at stage $j+1$ is a function of the transmission rate at stage $j$ and the intervention strategy $i$ taken at stage $j$. Equation \eqref{decision} indicates that only one intervention measure can be taken at each stage $j$. Equation \eqref{binary} describes the binary nature of intervention decisions. 

The transmission rate in our model is not equal to the basic reproduction number, $R_0$. It shows how many new tested infections will be caused by the symptomatic and asymptomatic infections from the previous stage. Since the number of new asymptomatic infections is uncertain, the number of new infections (both symptomatic and asymptomatic) changes under different scenarios even if the transmission rates at each stage $j$ stay the same.

There is a delay in the impact of the government' interventions on the number of infections and the reaction to the test results is also slow. Therefore, we calculate the transmission rate for the first two stages directly using the real data from \cite{JHU}, independent from the intervention type. Based on the first two-stage transmission rates, we calculate the transmission rates from stages three to five using the formulation \eqref{timerate}--\eqref{binary} for each intervention strategy.  Also, the values of $m_{j,t}^i$ are trained using the real data obtained from \cite{JHU}. As an example, the initial transmission rates for the first two stages in New York and New Jersey and the impacts of government intervention strategies $m_{j,t}^i$ are shown in Table \ref{Transmission} under Section \ref{EpidemiogicalData}.

\subsection{Uncertainty and Multi-period Scenario Tree}
Data regarding undetected or untested asymptomatic cases is lacking and uncertain. Therefore, we model the uncertainty regarding the proportion of untested asymptomatic infections $(\sigma_{2,r}^{\omega})$  by generating a set of scenarios $\omega \in \Omega$, each representing a specific realization of the uncertain proportion of untested asymptomatic individuals over multiple time periods. Our scenario generation approach is similar to \cite{alonso2018risk}'s method developed to model the demand uncertainty in forestry management. Each scenario has a probability of $p^{\omega}$ and $\sum\limits_{\omega \in \Omega}p^{\omega}=1$. Since data is not available to describe the probability distribution of the uncertain variable $(\sigma_{2,r}^{\omega})$, we assume that the uncertain parameter follows a normal distribution. The lower and upper bounds for the proportion of asymptomatic infections are obtained from the study of \cite{proportion}. The lower bound value for the random variable is considered as the value of 0.001-quantile and the upper bound is considered as the value of 0.999-quantile of the normal distribution.

\begin{figure}[h!]
\centering
\includegraphics[width=6in]{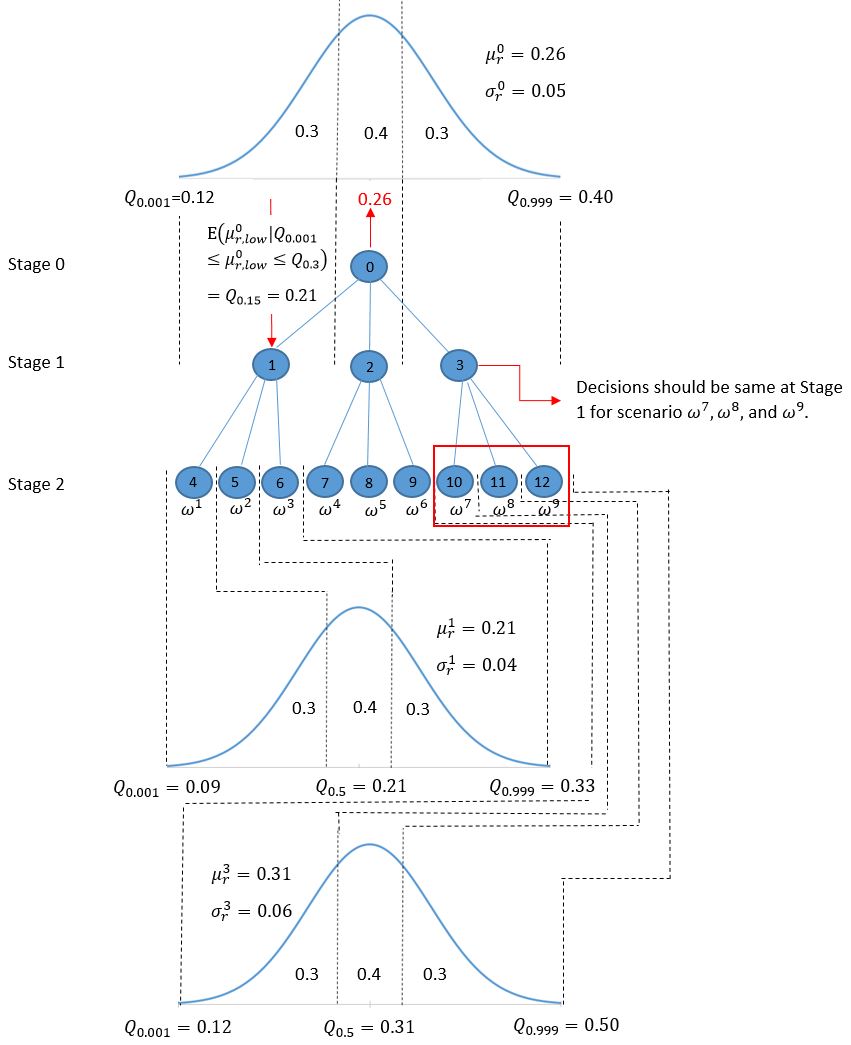}
\caption{Multi-stage scenario tree generation example for the uncertain proportion of untested asymptomatic infections ($\sigma^{\omega}_{2,r}$).}
\label{AsymptomaticScenario}
\end{figure}

As an example, Figure \ref{AsymptomaticScenario} shows a particular scenario tree for the proportion of untested asymptomatic infections $(\sigma_{2,r}^{\omega})$ for a two-stage problem. We consider three realizations at each node of the scenario tree by dividing its normal distribution into three discrete parts [low (L), medium (M), high (H)]. The low and high realizations have a probability of 0.3, and the medium realization has a probability of 0.4. Each path from the root node to the leaf node of the scenario tree represents a scenario $\omega$. The probability of a scenario $\omega$, $p^{\omega}$, is calculated as the multiplication of probabilities on the path for scenario $\omega$. For two stages, 9 ($3^2$) scenarios will be generated in this instance.  The non-anticipativity constraints indicate that two scenarios are inseparable at a stage $j$ if they share the same scenario path up to that stage. This means that the corresponding decision made at this stage for those two scenarios should also be the same.  

The value of the proportion of asymptomatic infections has a mean $\mu^j_r$ and standard deviation $\sigma^j_r$ at stage $j$. We use $Q_h$ to represent the value of $h$-quantile in the normal distribution. For each node $n$ in the scenario tree, the mean value of the low realization is the value of 0.15-quantile $(E(\mu^n_{r,low}|Q_{0.001}\leq \mu^n_{r,low} \leq Q_{0.30}) = Q_{0.15})$, the mean value of medium realization is the value of 0.50-quantile $(E(\mu^n_{r,medium}|Q_{0.30} \leq \mu^n_{r,medium} \leq Q_{0.70}) = Q_{0.50})$, and the mean value of high realization is the value of 0.85-quantile $(E(\mu^n_{r,high}|Q_{0.70} \leq \mu^n_{r,high} \leq Q_{0.999}) = Q_{0.85})$. For node $0$ in our example, the proportion of untested asymptomatic infections at stage $j = 0$ has $\mu^0_r$ = 0.26 and $\sigma_r^0$ = 0.05. The low, medium, and high realizations at node $0$ in stage $j = 0$ and nodes $1$ and $3$ in stage $j = 1$ are given in Table \ref{ScenarioExample} below. According to the distributions presented in Table \ref{ScenarioExample}, the proportion of untested asymptomatic infections in stage $1$ is realized as 0.21 (Low) at node 1, 0.26 (Medium) at node 2, and 0.31 (High) at node 3.

\begin{table}[h!]
  \centering
  \caption{The 0.15-, 0.50-, 0.85-quantiles of the normal distribution at nodes 0, 1, and 3 of the scenario tree in Figure \ref{AsymptomaticScenario} and the associated node of the uncertain parameter realization.}
    \begin{tabular}{ccccccccc}
    \hline
    & Low (realized node) & Medium (realized node) & High (realized node)\\
    & $Q_{0.15}$ & $Q_{0.50}$ & $Q_{0.85}$\\
    \hline
    Node 0 Distribution & 0.21 (node 1) & 0.26 (node 2) & 0.31 (node 3)\\
    Node 1 Distribution & 0.17 (node 4) & 0.21 (node 5) & 0.25 (node 6)\\
    Node 3 Distribution & 0.12 (node 10) & 0.31 (node 11) & 0.50 (node 12)\\
    \hline
    \end{tabular}%
  \label{ScenarioExample}%
\end{table}%

\subsection{Model Features and Assumptions}
Since the transmission of COVID-19 is affected by many factors, data to calibrate some of the model parameters, such as the impact of human mobility, is either lacking or inaccurate. Therefore, we incorporate some important features and make some assumptions in the model formulation. \\

\noindent \textbf{Important features.} First, we consider the impact of different intervention strategies on the disease transmission rate and adjust the short-term migration population depending on the intervention strategy. For instance, under the lockdown strategy, we assume that the short-term migration among each county is zero. Under mask and social distancing strategies, the short-term migration population among each county is reduced to 60\% of the original value, as estimated from the study of \cite{lee2020human}. Second, we incorporate the cost for purchasing ventilators to provide a capacity limitation on the total number of ventilators that could be allocated for treating COVID-19 patients. Since there are significant fluctuations in the ventilator prices \citep{VentilatorCost}, we consider the minimum purchase price for each ventilator acquired. Third, we train the real data to determine the impact rate of each intervention strategy on the disease transmission rate. The trained value of the impact of interventions can only be used in the regions considered in our case study since all the selected counties in New York and New Jersey are geographically close to each other, and thus interventions have similar social effects. However, for example, the impact of intervention strategies in a rural area may be different from those taken in a city. To estimate the COVID-19 transmission in other regions of the United States, the model should be re-trained using the associated data. \\

\noindent \textbf{Assumptions.} First, since studied counties in New Jersey and New York are geographically close to each other, the proportions of untested asymptomatic infections at each stage $j$ under scenario $\omega$ are set to be the same for each region $r$. Second, the model considers allocating newly purchased ventilators for the treatment of COVID-19 patients instead of re-allocating existing ventilators from other counties or states since the demand for ventilators during the disease's peak periods is high for all the counties and states, and there is a lead time for transfer of the ventilators between the states that are far from each other. Here, we also assume a central decision maker entitled to allocate a given supply of ventilators to multiple regions. Third, the infected individuals who cannot be treated in the hospital (both severe and less severe) due to the limited capacity have the same death rate as the ICU patients because some of those infections may worsen without professional treatment. Fourth, we assume that all symptomatic individuals are tested, and asymptomatic infected individuals are untested. Fifth, we assume that people react to the pandemic by anticipating the government's interventions and may start social distancing and quarantining days or weeks before an intervention is imposed \citep{zhang2020effects,fischer2020behavioural}. Thus, the transmission rate with either doing nothing or mask and social distancing shows a decreasing trend in later stages of the pandemic due to physical distancing among people. Lastly, we use each county's ICU capacity from \cite{JHU} as the initial ventilator availability.  We assume that non-COVID-19 patients use 60\% of this capacity \citep{mehrotra2020model}. Thus, only 40\% of the initial ICUs are available for treating the COVID-19 patients.

\subsection{Multi-Stage Risk and Time Consistency}
\label{Mean-Risk Problem}
The $\alpha$-quantile of the cumulative distribution of a random variable $z$, $\inf_{\eta} \lbrace \eta \in \mathbb{R} : F_z(\eta) \geq \alpha \rbrace$, is defined as the value-at-risk (VaR) at the confidence level $\alpha \in \lbrace 0, 1 \rbrace$ and denoted by $\textrm{VaR}_{\alpha}(z)$. The conditional expected value that exceeds the VaR at the confidence level $\alpha$ is called conditional value-at-risk (CVaR), defined as $\textrm{CVaR}_\alpha(z) = \mathbb{E}(z \mid z \geq \textrm{VaR}_\alpha(z))$. For a minimization problem, $\textrm{VaR}_\alpha$ is the $\alpha$-quantile of the distribution of the cost, and it provides an upper bound on the cost that is exceeded only with a small probability of $1-\alpha$. $\textrm{CVaR}_\alpha$ measures an expectation of the cost that is more than $\textrm{VaR}_\alpha$, and can be calculated as an optimization problem as follows \citep{rockafellar2002conditional}:
$$\textrm{CVaR}_\alpha(z)=\inf \limits_{\eta \in \mathbb{R}}\lbrace \eta + \frac{1}{1-\alpha}\mathbb{E}([z-\eta]_+)\rbrace,$$
where $(a)_+ := max(a,0)$ for any $a \in  \mathbb{R}$.

We formulate our model as a mean-risk minimization problem:
\begin{equation} \label{eq:mean_risk}
\min\limits_{x \in X} \lbrace \mathbb{E}(f(x, \omega))+\lambda \textrm{CVaR}_\alpha (f(x, \omega)) \rbrace,
\end{equation}
where $\mathbb{E}(f(x, \omega))$ is the expected cost function over the scenarios $\omega \in \Omega$, $\textrm{CVaR}_\alpha$ represents the conditional value-at-risk at $\alpha$, and $\lambda \in [0,1]$ is a non-negative weighted risk coefficient and it can be adjusted for a trade-off between optimizing an expectation value and the level of risk taken. \\

\noindent \textbf{Time Consistency.} Time consistency is considered as a critical issue when modeling a risk-averse multi-stage stochastic program. Time consistency means that if you solve a multi-stage stochastic programming model today, 
you should get the same solution if you resolve the problem tomorrow given the information that is observed and decided today. We consider a nested risk measure, expected conditional value-at-risk ($\mathbb{E}\textrm{-CVaR}$), as defined in \cite{homem2016risk} since it is shown to satisfy the time consistency of multi-stage stochastic programs. The $\mathbb{E}\textrm{-CVaR}$ can be linearized and formulated as a linear stochastic programming model. In the following section, we will utilize the $\mathbb{E}\textrm{-CVaR}$ as a risk measure in our formulation.

\subsection{Mathematical Model Formulation and Description}

The mathematical formulation for our risk-averse multi-stage stochastic epidemics-ventilator-logistics model is given below.\\

\noindent\textbf{Epidemics-Ventilator-Logistics Model Formulation:}

\begin{subequations}
\label{COVID:1}
\begin{eqnarray}
\min   && \sum\limits_{j\in J}\sum\limits_{\omega\in \Omega}p^{\omega}\left(\sum\limits_{r\in R}(I_{j,r}^{\omega}+F_{j,r}^{\omega})+\lambda(\eta_j^\omega+\frac{1}{1-\alpha}z_j^\omega) \right)\label{objective1-ex:1}\\
\textrm{s.t.} &&   S_{j+1,r}^{\omega} = S_{j,r}^{\omega}-\sigma_{1,r}(I_{j,r}^{\omega}+X_{j,r}^{\omega})-\sigma_{1,r}(I_{j,r}^{\omega}+X_{j,r}^{\omega})\frac{\sigma_{2,r}^{\omega}}{1-\sigma_{2,r}^{\omega}} \nonumber\\    
&&	j\in J\setminus \lbrace \overline{J} \rbrace, r\in R, \forall\omega\in \Omega, \label{cons1-ex:1}\\
&& I_{(j+1),r}^{\omega}=I_{j,r}^{\omega}+\breve{I}_{1,j,r}^\omega+\sigma_{1,r}(I_{j,r}^{\omega}+X_{j,r}^{\omega})-\lambda_1I_{j,r}^{\omega}-\lambda_2\overline{I}_{j,r}^{\omega}-O_{j,r}^{\omega} \nonumber\\
&&  j\in J\setminus \lbrace \overline{J} \rbrace, r\in R, \forall\omega\in \Omega, \label{cons1-ex:3}\\
&& X_{(j+1),r}^{\omega}=X_{j,r}^{\omega}+\sigma_{1,r}(I_{j,r}^{\omega}+X_{j,r}^{\omega})\frac{\sigma_{2,r}^{\omega}}{1-\sigma_{2,r}^{\omega}}-\lambda_{9}X_{j,r}^{\omega}\nonumber\\
&&  j\in J\setminus \lbrace \overline{J} \rbrace, r\in R, \forall\omega\in \Omega, \label{cons1-ex:4}\\
&& H_{(j+1),r}^{\omega}=H_{j,r}^{\omega}+O_{j,r}^{\omega}-\lambda_4H_{j,r}^{\omega}-\lambda_{5}K_{j,r}^{\omega}-\overline{C}_{j,r}^{\omega} \nonumber\\
&& j\in J\setminus \lbrace \overline{J} \rbrace, r\in R, \forall\omega\in \Omega, \label{cons1-ex:6}\\
&& C_{(j+1),r}^{\omega}=C_{j,r}^{\omega}+\overline{C}_{j,r}^{\omega}-\lambda_7C_{j,r}^{\omega}-\lambda_8C_{j,r}^{\omega} \qquad j\in J\setminus \lbrace \overline{J} \rbrace, r\in R, \forall\omega\in \Omega, \label{cons1-ex:7}\\
&& R_{(j+1),r}^{\omega}=R_{j,r}^{\omega}+\lambda_1I_{j,r}^{\omega}+\lambda_{9}X_{j,r}^{\omega}+\lambda_4H_{j,r}^{\omega}+\lambda_7C_{j,r}^{\omega} \nonumber \\
&& j\in J\setminus \lbrace \overline{J} \rbrace, r\in R, \forall\omega\in \Omega, \label{cons1-ex:12}\\
&& F_{(j+1),r}^{\omega}=F_{j,r}^{\omega}+\lambda_2\overline{I}_{j,r}^{\omega}+\lambda_5K_{j,r}^{\omega}+\lambda_{8}C_{j,r}^{\omega} \nonumber\\
&& j\in J\setminus \lbrace \overline{J} \rbrace, r\in R, \forall\omega\in \Omega, \label{cons1-ex:13}\\
&& O_{j,r}^{\omega}=min\lbrace \lambda_{3,r}I_{j,r}^{\omega}, T_{j,r}^{\omega}-H_{j,r}^{\omega} \rbrace \qquad j\in J, r\in R, \forall\omega\in \Omega, \label{cons1-ex:21}\\
&& \overline{C}_{j,r}^{\omega}=min\lbrace \lambda_{6,r}H_{j,r}^{\omega}, U_{j,r}^{\omega}-C_{j,r}^{\omega} \rbrace \qquad j\in J, r\in R, \forall\omega\in \Omega, \label{cons1-ex:8}\\
&&   U_{j,r}^{\omega}=U_{0,r} + \sum\limits_{l=1}^{j}y_{l,r}^{\omega}, \qquad j\in J, r\in R,\forall\omega\in \Omega, \label{cons1-ex:9}\\
&& \overline{I}_{j,r}^{\omega} \geq \lambda_3I_{j,r}^{\omega}-(T_{j,r}^{\omega}-H_{j,r}^{\omega}) \qquad j\in J, r\in R,\forall\omega\in \Omega, \label{cons1-ex:22}\\
&& \overline{I}_{j,r}^{\omega} \geq 0 \qquad j\in J, r\in R,\forall\omega\in \Omega, \label{cons1-ex:23}\\
&& K_{j,r}^{\omega} \geq \lambda_6H_{j,r}^{\omega}-(U_{j,r}^{\omega}-C_{j,r}^{\omega}) \qquad j\in J, r\in R,\forall\omega\in \Omega, \label{cons1-ex:10}\\
&& K_{j,r}^{\omega} \geq 0 \qquad j\in J, r\in R,\forall\omega\in \Omega, \label{cons1-ex:11}\\
&& \sum\limits_{j\in J}\sum\limits_{r\in R}y_{j,r}^{\omega}e_1 \leq \Delta \qquad  \forall\omega\in \Omega, \label{cons1-ex:15}\\
&& z_j^\omega \geq \sum\limits_{r\in R}(I_{j,r}^{\omega}+F_{j,r}^{\omega})-\eta_j^\omega \qquad j\in J, \forall\omega\in \Omega, \label{cons1-ex:16}\\
&& z_j^\omega \geq 0 \qquad j\in J, r\in R, \forall\omega\in \Omega, \label{cons1-ex:17}\\
&&  y_{t(n),r}^{\omega}-y_{n,r}=0, \quad z_{t(n)}^{\omega}-z_{n}=0, \quad \eta_{t(n)}^{\omega}-\eta_{n}=0, \qquad \forall\omega \in \beta(n), \forall n \in N, \label{cons1-ex:18}\\
&&   S_{j,r}^{\omega}, \quad I_{j,r}^{\omega}, \quad T_{j,r}^{\omega}, \quad H_{j,r}^{\omega}, \quad C_{j,r}^{\omega}, \quad R_{j,r}^{\omega}, \quad F_{j,r}^{\omega}, \quad B_{j,r}^{\omega}, \quad \overline{C}_{j,r}^{\omega}, \quad O_{j,r}^{\omega} \geq 0, \nonumber \\
       &&   j\in J, r\in R, \forall\omega\in \Omega, \label{cons1-ex:19}\\
       &&   y_{j,r}^{\omega}\in \lbrace 0, 1, 2, \ldots,  \frac{\Delta}{e_{1}}\rbrace \qquad j\in J\setminus \lbrace \overline{J} \rbrace, r\in R, \forall\omega\in \Omega. \label{cons1-ex:20}
\end{eqnarray}
\end{subequations}

\noindent \textbf{Objective Function \eqref{objective1-ex:1}.} The objective function \eqref{objective1-ex:1} minimizes the total expected number of tested infected individuals and deaths and the conditional value-at-risk over all stages $j$ and scenarios $\omega$.\\

\noindent \textbf{Population Infection Dynamics Constraints \eqref{cons1-ex:1} - \eqref{cons1-ex:13}.}  Constraint \eqref{cons1-ex:1} represents that the number of susceptible individuals in region $r$ at stage $j+1$ under scenario $\omega$ equals the number of susceptible individuals at stage $j$ minus the number of susceptible individuals who become either tested infected or untested asymptomatic infected at stage $j$. In this equation, the number of untested asymptomatic infections equals the number of tested infections multiplied by the proportion of the untested asymptomatic infections to the tested infections. Constraint \eqref{cons1-ex:3} shows that the number of tested infected individuals in region $r$ at stage $j+1$ under scenario $\omega$ equals the number of tested infected individuals at stage $j$ plus the infected individuals caused by short-term migration, plus the newly tested infections at time $j$, minus the recovered and deceased infections of tested individuals at stage $j$, minus the hospitalized individuals at stage $j$. Constraint \eqref{cons1-ex:4} implies that the number of untested asymptomatic infections in region $r$ at stage $j+1$ under scenario $\omega$ equals the number of untested asymptomatic infections at stage $j$ plus new, untested asymptomatic infections, minus the recovered untested asymptomatic infections at stage $j$. Constraint \eqref{cons1-ex:6} shows that the hospitalized individuals in region $r$ at stage $j+1$ under scenario $\omega$ equals the number of hospitalized individuals at stage $j$ plus the newly hospitalized individuals at stage $j$, minus the recovered and deceased individuals at stage $j$, minus the individuals who move to the intensive care unit (ICU) at stage $j$. Constraint \eqref{cons1-ex:7} indicates that the total number of individuals in ICU in region $r$ at stage $j+1$ under scenario $\omega$ equals the total number of individuals in ICU at stage $j$ plus the individuals who moved to ICU at stage $j$, minus the individuals who are recovered or died at the ICU at stage $j$. Constraint \eqref{cons1-ex:12} shows that the number of recovered individuals in region $r$ at stage $j$ under scenario $\omega$ equals the number of recovered individuals at stage $j$ plus the recovered individuals from tested infected, untested asymptomatic infected and hospitalized individuals, and ICU patients at stage $j$. Constraint \eqref{cons1-ex:13} indicates that the number of deceased individuals in region $r$ at stage $j+1$ under scenario $\omega$ equals the number of deceased individuals at stage $j$ plus the deceased individuals from tested infected and hospitalized individuals and ICU patients at stage $j$.\\

\noindent \textbf{Ventilator Logistics and Capacity Constraints \eqref{cons1-ex:21} - \eqref{cons1-ex:15}.} Constraint \eqref{cons1-ex:21} ensures that the number of individuals admitted to the hospital in region $r$ at stage $j$ under scenario $\omega$ equals the minimum number of individuals who require hospitalization and the available hospital capacity at stage $j$. Constraint \eqref{cons1-ex:8} implies that the number of individuals admitted to ICU in region $r$ at stage $j$ under scenario $\omega$ equals the minimum number of individuals who require treatment in ICU and the number of available ventilators at stage $j$. Constraint \eqref{cons1-ex:9} represents that the cumulative number of ICU beds (equivalent to ventilators) in region $r$ at stage $j$ under scenario $\omega$ equals the initial number of ICU beds plus the cumulative number of ICU beds (new ventilators) allocated from stage 1 to stage $j$. Constraints \eqref{cons1-ex:22} - \eqref{cons1-ex:11} show that the number of individuals who can not be admitted to the hospital or the ICU due to limited capacity should be greater than or equal to zero. Constraint \eqref{cons1-ex:15} represents that the cost of ventilators allocated over all regions and time stages under scenario $\omega$ cannot exceed the total budget allocated for ventilators. The budget here also represents the maximum total ventilator supply that could be available throughout the planning horizon.\\

\noindent \textbf{Risk Measure Constraints  \eqref{cons1-ex:16} and \eqref{cons1-ex:17}.} Constraint \eqref{cons1-ex:16} indicates the difference between the objective function value and the value-at-risk for each stage $j$ under each scenario $\omega$. Constraint \eqref{cons1-ex:17} ensures that the loss value exceeding the value-at-risk is included in the CVaR calculation, and thus $z_j^\omega$ should be greater than or equal to zero.\\

\noindent \textbf{Non-Anticipativity, Non-Negativity and Integrality Constraints \eqref{cons1-ex:18} - \eqref{cons1-ex:20}.} Constraint \eqref{cons1-ex:18} is the non-anticipativity constraint, indicating that the scenarios that share the same path up to stage $j$ should also have the same corresponding decisions. Constraint \eqref{cons1-ex:19} indicates that the number of individuals in each compartment in region $r$ at stage $j$ under scenario $\omega$ should be greater or equal to zero. Constraint \eqref{cons1-ex:20} implies that the number of allocated ventilators should be an integer.\\

\begin{remark} Both \eqref{cons1-ex:21} and \eqref{cons1-ex:8} are non-linear, and thus we replace them with equivalent linear constraints with additional linearization variables, using the method presented in \cite{yinbuyuktahtakin2020}. Hence, the non-linear multi-stage stochastic programming epidemics-ventilator-logistics model \eqref{objective1-ex:1}--\eqref{cons1-ex:20} is converted into an equivalent mixed-integer linear programming (MIP) formulation. We implement this MIP formulation for the rest of the paper.
\end{remark}

\subsection{Scenario Sub-Problem and Bounds}\label{Cuts}
We implement the scenario sub-problems and lower and upper bounds proposed by \cite{und_rev_3} to reduce the optimality gap of solving our risk-averse multi-stage stochastic programming problem \eqref{objective1-ex:1}--\eqref{cons1-ex:20}, while referring to \cite{und_rev_3} for the proofs of those bounds originally driven for the general multi-stage stochastic programs. The scenario sub-problem and bounds are described below.
 
\begin{definition} The scenario-$\omega$ problem $(P^\omega)$ is formulated as follows:
\begin{subequations}
\label{Scen-w}
\begin{eqnarray}
   Z^\omega = \min   &&  \sum\limits_{j\in J}p^{\omega}\left(\sum\limits_{r\in R}(I_{j,r}^{\omega}+F_{j,r}^{\omega})+\lambda(\eta_j^\omega+\frac{1}{1-\alpha}z_j^\omega) \right)
  \label{sd3} \\
\textrm{s.t.}        &&  \text{Constraints} \quad \eqref{cons1-ex:1}-\eqref{cons1-ex:20}. \label{sd4} 
\end{eqnarray}
\end{subequations}
\end{definition}
Specifically, in $P^\omega$ we minimize the objective function value only under scenario $\omega$ while keeping all the variables and the constraints from the original problem  \eqref{objective1-ex:1}--\eqref{cons1-ex:20}.

\begin{proposition} (\textbf{Lower Bound}) Let $Z^*$ represent the objective function of the original problem \eqref{objective1-ex:1}--\eqref{cons1-ex:20}, $P$. Then we have:
\begin{align}
Z^* \geq \sum\limits_{\omega \in \Omega}Z^\omega. \label{sd5}
\end{align}
\end{proposition}


\begin{proposition} (\textbf{Upper Bound}) Let $\dot{x}^\omega$ be the optimal solution of scenario-$\omega$ problem, $P^\omega$, and $Z(\dot{x}^\omega)$ be the objective value of original problem \eqref{objective1-ex:1}--\eqref{cons1-ex:20} 
where $\dot{x}^\omega$ is substituted in the original problem objective function \eqref{objective1-ex:1}. 
Then we have:
\begin{align}
Z^* \leq \min\limits_{\omega \in \Omega}Z(\dot{x}^\omega). \label{sd6}
\end{align}
\end{proposition}

\section{Case Study Data}
\label{Case Study Data}
\begin{figure}[h!]
\centering
\includegraphics[width=3in]{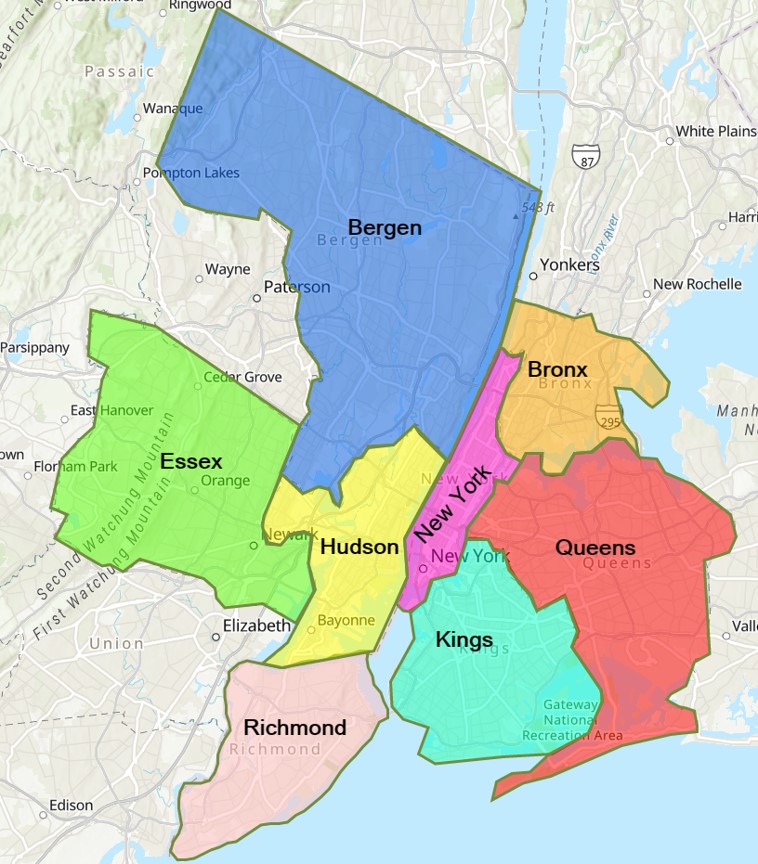}
\caption{Counties in New York and New Jersey (Source: \cite{ArcGIS})}
\label{Region}
\end{figure}

This section provides the data used to calibrate model parameters and formulate the model, including population and short-term migration data, transmission parameters, as well as the cost of a ventilator. As shown in Figure \ref{Region}, we select eight counties that are most impacted by the pandemic in the states of New York and New Jersey for our case study. They are New York County, Kings County, Queens County, Bronx County, Richmond County, Hudson County, Bergen County, and Essex County. In our multi-stage model, each stage represents a two-week period. Thus, all the data regarding the transmission and migration are bi-weekly.

\subsection{Population and Migration Data}
Table \ref{populationdata} shows the population data for each considered county in New York and New Jersey. Population data is obtained from \cite{JHU}. The migration rates among the considered counties, estimated from the data on \cite{CENSUS}, are presented in Table \ref{MigrationRate}. The blank areas in Table \ref{MigrationRate} represent a zero short-term migration because the movement among those counties is too small to make an impact on the model results.
\begin{table}[H]\footnotesize
  \centering
	\renewcommand\arraystretch{0.9}
\renewcommand\tabcolsep{3.5pt}
  \caption{Counties and population sizes in New York and New Jersey}
    \begin{tabular}{llllll}
    \hline
    New York&  & Population & New Jersey & Population\\
    \hline
    New York & &  1,632,480  &  Hudson & 668,631\\
    Kings   & & 2,600,747 &  Bergen & 929,999\\
    Queens  & & 2,298,513 &  Essex & 793,555\\
    Bronx & & 1,437,872 &\\
    Richmond & & 474,101 &\\
    \textbf{Total} & & \textbf{8,443,713 } & & \textbf{2,392,185}\\
    \hline
    \end{tabular}%
  \label{populationdata}%
\end{table}%

\begin{table}[H]\footnotesize
  \centering
	\renewcommand\arraystretch{0.9}
\renewcommand\tabcolsep{3.0pt}
  \caption{Migration rate among counties in New York and New Jersey}
    \begin{tabular}{lcccccccc}
    \hline
    To  &  New York & Kings & Queens & Bronx & Richmond & Hudson & Bergen & Essex\\
    From &  &  & & & & &\\
    \hline
    New York & & 0.015 &0.012&0.009& 0.006 &0.007&0.007&0.007\\
    Kings   & 0.192& &0.038&0.004& 0.004 & & &\\
    Queens  & 0.218& 0.044 & &0.009& 0.002 & & &\\
    Bronx & 0.209& 0.014 &0.028& & &0.003& &0.003\\
    Richmond & 0.105& 0.105 & & & & & &\\
    Hudson &0.040& & & & 0.001 & &0.040&0.040\\
    Bergen &0.126& & & & &0.039& &0.039\\
    Essex &0.079& & & & 0.001 &0.057& 0.057 &\\
    \hline
    \end{tabular}%
  \label{MigrationRate}%
\end{table}%

\subsection{Epidemiological Data}\label{EpidemiogicalData}

Table \ref{epidemicdata} presents the data values for transmission parameters for the studied counties in New York and New Jersey. The data contains the proportion of untested asymptomatic infections, recovery rate, and the death rate for tested infections, hospitalized infections, and ICU patients. Table \ref{Transmission} shows the transmission rate of each county at the first two stages and the impacts of applying different intervention strategies, as discussed in Section \ref{TimeRate}.

\begin{table}[H]\footnotesize
  \centering
		\renewcommand\arraystretch{0.9}
\renewcommand\tabcolsep{3.0pt}
  \caption{Transmission parameters and bi-weekly rates for the COVID-19}
    \begin{tabular}{llllll}
    \hline
    Parameter & Description & Data & & Reference \\
  & & NY & NJ & & \\
    \hline
    $\sigma_{2,r}$   & Proportion of untested asymptomatic infections & 0.15-0.4 & 0.15-0.4 & \cite{proportion}\\
    $\lambda_{1}$   & Recovery rate without hospitalization & 0.69-0.79 & 0.69-0.79 & \cite{hospitalrate}\\
    $\lambda_{2}$   & Death rate without hospitalization & 0.4 & 0.4 & Trained using real data \citep{JHU}\\
    $\lambda_{3}$   & Hospitalization rate & 0.21-0.31 & 0.21-0.31 &  \cite{hospitalrate}\\
    $\lambda_{4}$   & Recovery rate with hospitalization & 0.88 & 0.88 & \cite{hospitalrate}\\
    $\lambda_{5}$   & Death rate with hospitalization (No ventilators) & 0.4 & 0.4 & Trained using real data \citep{JHU}\\
    $\lambda_{6}$   & Ventilator requirement rate of hospitalized & 0.12 & 0.12 & \cite{hospitalrate}\\
    $\lambda_{7}$   & Recovery rate with ventilator & 0.643 & 0.643 & \cite{ventilatordeath}\\
    $\lambda_{8}$   & Death rate with ventilator & 0.357 & 0.357 & \cite{ventilatordeath}\\
		$\lambda_{9}$   & Recovery rate with asymptomatic infections & 1 & 1 & \cite{bertsimas2020predictions}\\
    \hline
    \end{tabular}%
  \label{epidemicdata}%
\end{table}%

\begin{table}[H]\footnotesize
  \centering
	\renewcommand\arraystretch{0.9}
\renewcommand\tabcolsep{3.0pt}
  \caption{Transmission rate $(\sigma_{1,r})$ in New York and New Jersey and impact of interventions}
    \begin{tabular}{lccccccccc}
    \hline
    County  &  Transmission Rate & Transmission Rate & Impact of & Impact of & Impact of \\
    & at Stage 1 & at Stage 2 & None & Mask and Social Distancing & Lockdown\\
    \hline
    New York & 4.5 & 0.9855 & 1 & 0.4 & 0.6\\
    Kings & 9 & 0.9855 & 1 & 0.4 & 0.6 \\
    Queens & 10 & 1.095 & 1 & 0.4 & 0.6\\
    Bronx & 12 & 1.314 & 1 & 0.4 & 0.6\\
    Richmond & 12 & 1.314 & 1 & 0.4 & 0.6\\
    Hudson & 22 & 2.409 & 1 & 0.3 & 0.6\\
    Bergen & 11 & 1.408 & 1 & 0.3 & 0.6\\
    Essex & 22 & 2.409 & 1 & 0.3 & 0.6\\
    \hline
    \end{tabular}%
  \label{Transmission}%
\end{table}%

\subsection{Initial Infection, Capacity and Cost Data}

Table \ref{Initial Data} shows the initial number of infections, hospital capacity, and ICU capacity for each county. The data is obtained from \cite{JHU}.

\begin{table}[H]\footnotesize
  \centering
	\renewcommand\arraystretch{0.9}
\renewcommand\tabcolsep{3.0pt}
  \caption{Initial number of infections, hospital capacity, and ICU capacity for each county}
    \begin{tabular}{lccccccccc}
    \hline
    County  &  Initial & Initial & Initial \\
    & Infections & Hospital Capacity & ICU Capacity\\
    \hline
    New York & 1200& 8650& 944\\
    Kings & 1300& 5838&282\\
    Queens & 1100& 3210&146\\
    Bronx & 554& 2816&274\\
    Richmond & 206& 1177&72\\
    Hudson & 66& 1764&89\\
    Bergen & 249& 2874&122\\
    Essex & 73& 3541&226\\
    \hline
    \end{tabular}%
  \label{Initial Data}%
\end{table}%

\noindent\textbf{Ventilator Cost.} The cost of each ventilator ranges from \$5000 to \$50000 \citep{VentilatorCost}. In our case, we consider a cost of \$5000 for each ventilator, and different budget levels are set to impose different upper bounds on the ventilator supply.

\section{Results}

\subsection{Model Validation}
This section presents the validation results of the mathematical model in Eqs. \eqref{objective1-ex:1}--\eqref{cons1-ex:20} as presented in Section 3 for the 8-stage time period from April 3, 2020, to July 10, 2020. We consider a medium realization of the uncertain asymptomatic proportions at each stage of the planning horizon and compare the number of infections forecasted by our model to the real outbreak data.

The government applied a lockdown strategy from April 3, 2020, to July 10, 2020, at those considered locations in New York and New Jersey.  Thus, we use the lockdown strategy and the corresponding transmission rate at each stage in our model for validation.

\begin{figure}[h!]
\centering
\includegraphics[width=7in]{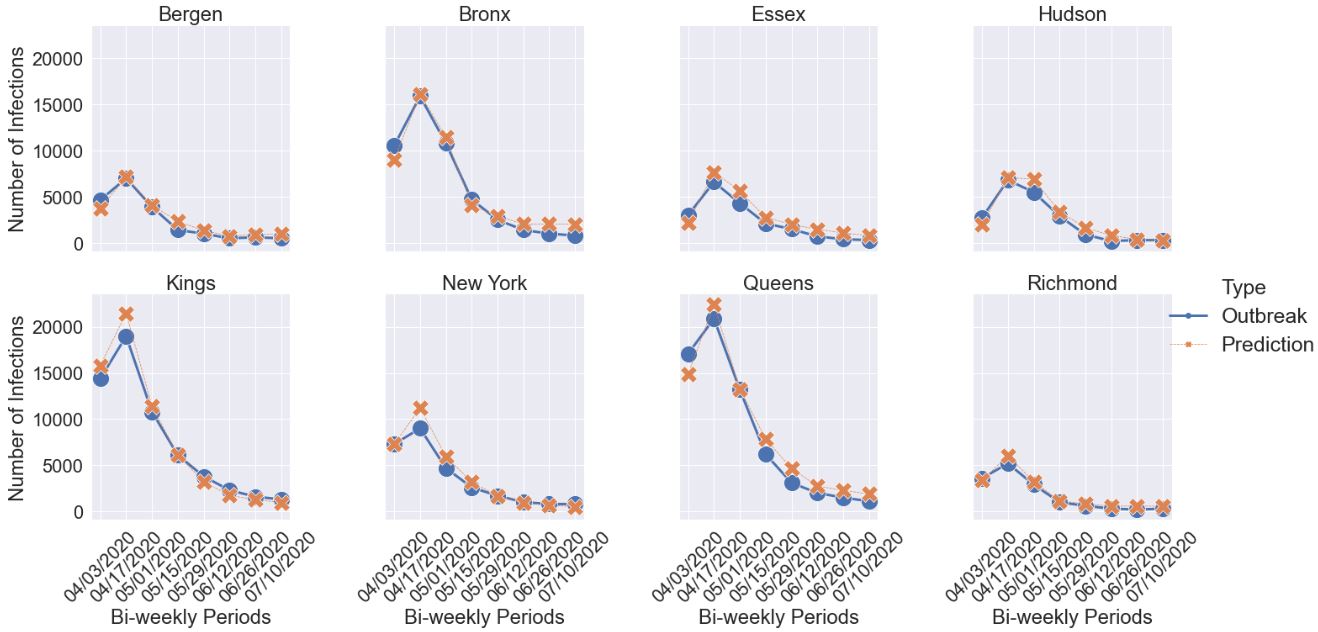}
\caption{Comparison of predicted cases with real outbreak data for new infections in New York and New Jersey}
\label{InfectionValidation}
\end{figure}

Figure \ref{InfectionValidation} shows the comparison between the predicted infections and real outbreak data. The model's predictions provide a visually good fit for the actual number of new infections in each region, implying that the model can capture the disease transmission dynamics under a lockdown intervention strategy. We also perform a paired-$t$-test to analyze the difference between the pairs of predicted new infections and the actual data in each period. As shown in Table \ref{Statistical Analysis}, all $p$-values are greater than 0.05, and thus our model provides statistically similar predictions with the real outbreak data from April 3, 2020, to July 10, 2020, for each considered county.

\begin{table}[H]
  \centering
  \caption{Statistical analysis to compare the bi-weekly predicted new cases and real outbreak data.}
    \begin{tabular}{llccccccc}
    \hline
     & County & \multicolumn{2}{c}{Mean} & \qquad & \multicolumn{3}{c}{Two-tailed paired-t-test} \\
     \cline{3-4}\cline{6-8}
     & & Outbreak & Predicted & & t-stat & t-critical & p-value \\
    \hline
      & New York & 7300 & 7299 & & 0.20 & 2.36 & 0.58\\
      & Kings & 7413 & 7754 & & 0.41 & & 0.65 \\
      & Queens & 8138 & 8751 & & 0.21 & & 0.58\\
      & Bronx & 5956 & 6214 & & 0.46 & & 0.67\\
      & Richmond & 1762 & 2040 & & 0.04 & & 0.51\\
      & Hudson & 2455 & 2806 & & 0.16 & & 0.56\\
      & Bergen & 2444 & 2656 & & 0.30 & & 0.61\\
      & Essex & 2366 & 2948 & & 0.04 & & 0.51\\
    \hline
    \end{tabular}%
  \label{Statistical Analysis}%
\end{table}%

\subsection{Case Study Implementation Details}
We apply our model described in Section \ref{Model} to the selected counties in New York and New Jersey. We first solve the risk-neutral model. We incorporate the uncertainty of the proportion of untested asymptomatic infections as well as the short-term migration in the disease transmission and forecast the number of new infections, deceased individuals, hospitalized individuals, and ICU patients under different intervention strategies. Also, we solve the model to determine the optimal location and number of ventilators allocated under different budget levels and scenarios to provide insights into resource allocation over multiple jurisdictions under uncertainty. Due to the high complexity of the mathematical formulation, we solve it for a 5-stage time period. Each stage corresponds to two weeks, resulting in a planning horizon of ten weeks from March 20, 2020, to May 29, 2020. Because each node of the scenario tree has three possible realizations of the random parameters, we solved 243 ($3^5$) scenarios simultaneously.

The mathematical model is solved using CPLEX 12.7.1 on a desktop computer running with Intel i7 CPU and 64.0 GB of memory. We set the time limit at 7200 CPU seconds to solve each test instance. We extend running time for specific budget levels ($\$30$ Million) and interventions (``Lockdown'') due to the large optimality gap. In the following subsections, we present results from solving the multi-stage stochastic epidemics-ventilator-logistics model with an application to the case of COVID-19 using the data presented in Section \ref{Case Study Data}.

\subsection{Transmission Forecast under Different Intervention Strategies}

Here, we present results of our formulation for each time period under different intervention strategies: No intervention (``None''), mask and social distancing for all stages (``Mask and Social Distance''), lockdown for all stages (``Lockdown''), mask and social distancing for the first three stages and lockdown for the following two stages (``Mask + Lockdown''), lockdown for the first three stages and mask and social distancing for the following two stages (``Lockdown + Mask''). The model is solved under the $ \$30 $ million budget level. The model with the ``Lockdown'' strategy gives a $4.54\%$ optimality gap after a run time of 43,241 CPU seconds, while the model solved for all other strategies has a zero optimality gap within 7,200 CPU seconds.

\begin{figure}[h!]
\centering
\includegraphics[width=7in]{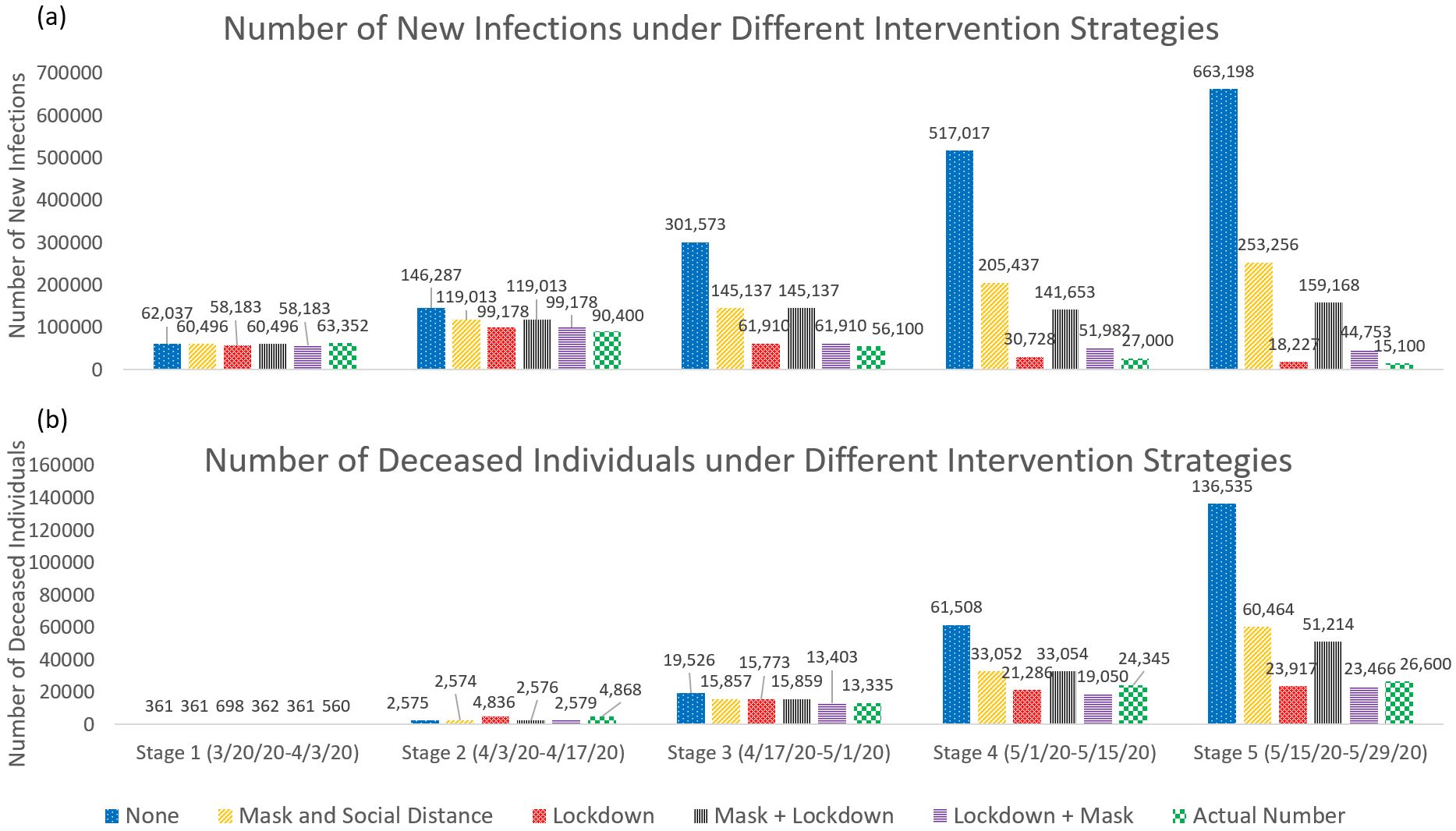}
\caption{Number of new infections and deaths under different intervention strategies and actual numbers}
\label{InterventionIF}
\end{figure}

Figure \ref{InterventionIF} presents the number of infections and deceased individuals at each stage under different intervention strategies. According to the results, short-term migration influences the number of new infections even under constant transmission rates. As in the first stage, within the same initial transmission rate, the number of infections under different intervention strategies is different from each other. When the stage increases, the difference in the number of new infections among each intervention strategy becomes more and more significant. The ``None'' strategy has the most infections at each stage, followed by the ``Mask and Social Distance'' strategy. The ``Lockdown'' strategy results in the lowest number of new infections compared to those under other strategies at each stage. The ``Lockdown'' strategy provides a little higher number of infections compared to the actual infection data since our model slightly (but statistically insignificantly) overestimates the number of infections. Compared to the ``Mask + Lockdown'' strategy, the ``Lockdown + Mask'' intervention leads to fewer infections. This implies that applying the ``Lockdown'' strategy immediately at the onset of the pandemic followed by the ``Mask and Social Distance'' intervention is a better strategy than enforcing ``Mask and Social Distance'' first and delaying the lockdown.

\begin{figure}[h!]
\centering
\includegraphics[width=7in]{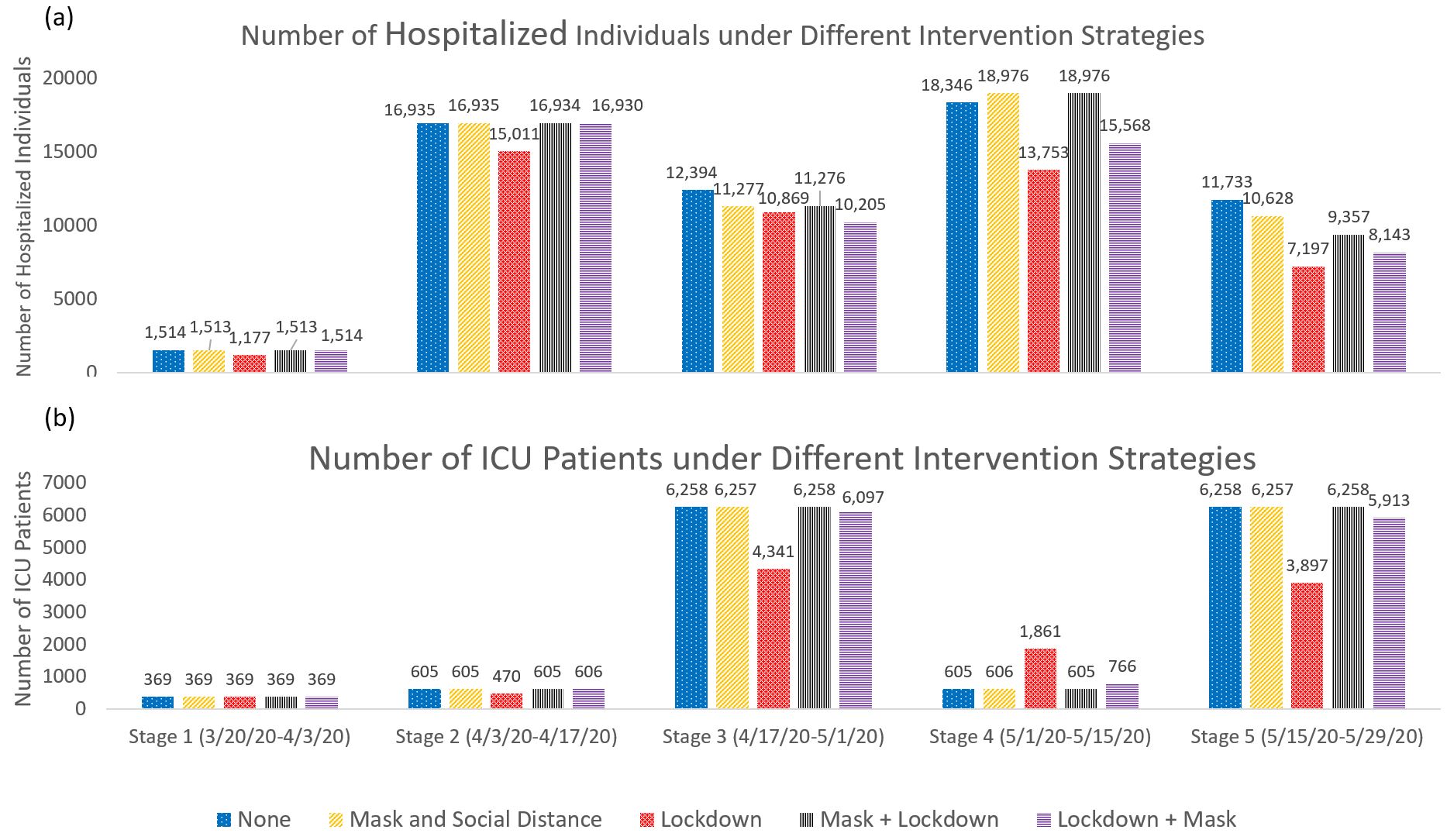}
\caption{Number of hospitalized individuals and ICU patients under different intervention strategies}
\label{InterventionHC}
\end{figure}

The intervention strategy does not influence the number of deceased individuals as quickly as it does impact the number of infections, as shown in Figure \ref{InterventionIF}. Here, the number of deceased individuals at the first two stages is much lower than that of the last three stages. Starting from stage three, the number of deceased individuals under different intervention strategies shows a similar trend with the number of new infections. The influence of interventions is further delayed for those confirmed infections to be treated in the hospital and ICU (Figure \ref{InterventionHC}). 

To reduce both the number of new infections and deaths, ``Lockdown'' is the best strategy.  As shown in Figure \ref{InterventionIF}b, the ``Lockdown'' strategy with the optimal ventilator allocation further reduces the actual number of deaths.  Due to the negative impact of COVID-19 on employment and its economic burden, governments are often forced to stop the lockdown and reopen businesses. In such cases, applying ``Mask and Social Distance'' after a certain period of ``Lockdown'' will be the best choice.

Figure \ref{InterventionHC} shows the number of hospitalized individuals and ICU patients at each stage under different intervention strategies. Similar to the number of deceased individuals, there are delays in the impact of government interventions on the number of hospitalized individuals and ICU patients. An infected person may have mild symptoms for about one week, then worsen rapidly (\cite{Worse}). Thus, it may take some time for patients to be admitted to the ICU, so the impact of interventions on the number of ICU patients is delayed one more stage compared to the hospitalized cases. As shown in Figure \ref{InterventionHC}, more number of hospitalized individuals at stage $j$ will lead to more ICU patients at stage $j + 1$.

For all the stages, the ``Lockdown'' strategy has the least number of hospitalized individuals and the ICU patients, followed by the ``Lockdown + Mask'' intervention. The ICU patients of ``None,'' ``Mask and Social Distance,'' and ``Mask + Lockdown'' are almost the same at stages three to five. This is because under those, the need for ventilators is large, and the number of treated individuals in ICUs depends on the minimum number of patients who require ventilators and the ventilator supply in those ICUs. So the number of treated patients in ICUs is limited by the tight ventilator availability.

\subsection{Optimal Ventilator Allocation}

Table \ref{AllocationResults} shows the number of ventilators allocated to each region at stages one and two and the total number of ventilators allocated throughout the planning horizon under different budget levels and three select scenarios. The ``All Low,'' ``All Medium,'' and ``All High'' scenarios represent low, medium, and high realization of the proportion of untested asymptomatic infections at each stage of a five-stage planning horizon, respectively. To analyze the impact of budget on the optimal ventilator allocation decisions, we select $ \$10$M as the limited budget level, $ \$20$M as the medium budget level, and $ \$30$M as the ample budget level. The model has zero optimality gap under the $ \$10$M budget level, $4.54\%$ optimality gap under the $ \$20$M budget level and $7.77\%$ optimality gap under the $ \$30$M budget level within two hours of solution time. 

The results in Table \ref{AllocationResults} demonstrate that the location and number of ventilators allocated depend on several factors, including the initial and evolving disease transmission rates, the population and the number of initial infections in a region, and the existing ventilator capacity. Thus, the optimal ventilator allocation should be determined case-by-case. 

\begin{table}[h!] \small
		\renewcommand\arraystretch{0.8}
\renewcommand\tabcolsep{3.0pt}
  \centering
  \caption{Optimal ventilators allocated under different scenarios for budget levels}
    \begin{tabular}{cc|ccc|ccc|ccccc}
    \hline
		   \textbf{Scenario} & \textbf{County} & \textbf{Stage} & \textbf{Stage} & \textbf{Total} & \textbf{Stage} & \textbf{Stage} & \textbf{Total}&\textbf{Stage} & \textbf{Stage} & \textbf{Total}\\
                  &      & \textbf{1} & \textbf{2}  & \textbf{Ventilator}& \textbf{1} & \textbf{2}  & \textbf{Ventilator}& \textbf{1} & \textbf{2}  & \textbf{Ventilator}\\
                   & &  \multicolumn{3}{|c|}{\textbf{(Budget=$\$10$M)}}  &   \multicolumn{3}{|c|}{\textbf{(Budget=$\$20$M)}}    & \multicolumn{3}{|c}{\textbf{(Budget=$\$30$M)}}   \\
    \hline
    \multirow{8}[1]{*}{All Low} & New York&0&0& 0&721&1&\boldmath{$802^{*}$} &2100&0&2100  \\
                 & Kings &119&0& 119&1734&0&\boldmath{$1736^{*}$} &2222&0&2222\\
                & Queens &107&1069 & 1176&0&0&0 &0&0&0\\
                 & Bronx &28&0 & 28&1016&1&1017 &1016&1&1017\\
                & Richmond &18&0& 18&0&0&0 &0&0&0\\
           & Hudson &0&250& 250 &0&0&0 &0&0&\boldmath{$2^{*}$}\\
                 & Bergen &187&0 & 187&12&433&445 &0&0&\boldmath{$165^{*}$}\\
                 & Essex &218&0& 218&0&0&0 &0&0&\boldmath{$174^{*}$}\\
    \textbf{Total}        &      &\textbf{677}&\textbf{1323} & \textbf{2000}&\textbf{3483}&\textbf{435} & \boldmath{$4000^{*}$}&\textbf{5338}&\textbf{1}& \boldmath{$5680^{*}$}\\
    \hline
    \multirow{8}[1]{*}{All Medium} & New York&0&0& 0  &721&0&721 &2100&0&\boldmath{$2101^{*}$}\\
                 & Kings&119&0& 119 &1734&76&1810 &2222&1&2223\\
                & Queens&107&0 & 107&0&0&\boldmath{$3^{*}$} &0&0&\boldmath{$69^{*}$}\\
                 & Bronx&28&809 & 837 &1016&0&1016 &1016&0&\boldmath{$1017^{*}$}\\
                 & Richmond&18&0& 18&0&0&0 &0&0&\boldmath{$35^{*}$}\\
           & Hudson &0&250& 250 &0&0&0 &0&0&\boldmath{$53^{*}$}\\
                 & Bergen&187&264 & 451 &12&435&447 &0&0&0\\
                  & Essex&218&0 & 218 &0&0&\boldmath{$3^{*}$} &0&0&\boldmath{$349^{*}$}\\
    \textbf{Total}        &      &\textbf{677}&\textbf{1323} & \textbf{2000}&\textbf{3483}&\textbf{511}& \boldmath{$4000^{*}$}&\textbf{5338}&\textbf{1}& \boldmath{$5847^{*}$}\\
    \hline
    \multirow{8}[1]{*}{All High}  & New York&0&0& 0 &721&0&721 &2100&1&\boldmath{$2190^{*}$} \\
                 & Kings&119&0& 119 &1734&0&1734 &2222&0&2222\\
                 & Queens&107&1055 & 1162&0&0&\boldmath{$2^{*}$} &0&0&\boldmath{$451^{*}$}\\
                 & Bronx&28&4& 32 &1016&0&1016 &1016&0&\boldmath{$1017^{*}$}\\
                 & Richmond&18&0& 18&0&0&\boldmath{$78^{*}$} &0&0&0\\
           & Hudson &0&0& 0 &0&0&0 &0&0&0\\
                 & Bergen&187&264& 451 &12&437&449 &0&0&\boldmath{$93^{*}$}\\
                  & Essex&218&0& 218 &0&0&0 &0&0&\boldmath{$27^{*}$}\\
    \textbf{Total}       &      &\textbf{677}&\textbf{1323}& \textbf{2000}&\textbf{3483}&\textbf{437}& \boldmath{$4000^{*}$}&\textbf{5338}&\textbf{1}& \boldmath{$6000^{*}$}\\
    \hline
    \end{tabular}%
    \begin{tablenotes}
        \footnotesize
        \item[*] \boldmath{$^{*}$} Some of the ventilators are allocated at stages three and four.
      \end{tablenotes}
  \label{AllocationResults}%
\end{table}%

According to the results, the total number of ventilators allocated increases in the budget level due to the high need for ventilators. As shown in Table  \ref{AllocationResults}, under the limited budget level, some regions with many initial infections, e.g., the New York County, do not receive ventilators. This situation is because the initial ventilator capacity of those regions is higher than in other counties. Also, results suggest that more ventilators should be distributed to other areas with a higher initial transmission rate than the New York County, such as Kings, Hudson, and Essex, under a very tight budget. Kings, Queens, and Hudson have higher initial transmission rates and lower initial ICU capacity than New York. Thus, these regions get more ventilators allocated under a limited budget level and ``All Low'' scenario. Also, regions with a relatively smaller population, such as Hudson and Essex in New Jersey, get a large share of ventilators with a very tight budget under the ``All Low'' and ``All Medium'' scenarios due to their high transmission rates at the beginning of the pandemic.

Independent from the budget level, some regions with low initial infections and low initial ICU capacity (e.g., Bronx) will get more ventilators allocated under the ``All Medium'' scenario. Under this scenario, the number of infections in regions with a high initial transmission rate (e.g., Kings and Queens) will not significantly increase even if they receive a smaller number of ventilators. These regions usually have much more initial ICU capacity for the treatment because of their large population. On the contrary, the areas with a lower initial transmission rate but less initial ICU capacity may benefit more if they receive more ventilators. As a comparison with the ``All Medium'' scenario, the number of infections in the region with a low initial transmission rate will be much smaller under the ``All Low'' scenario, and the number of infections in the regions with a high initial transmission rate will be much larger under the ``All High'' scenario. The model gives priority to allocate more ventilators to the regions with high initial transmission rates for both of the ``All Low'' and ``All High'' scenarios because the benefit of giving resources to those regions is higher than the regions with low initial disease transmission. 

Moreover, the model is forced to make difficult decisions, and some of the regions may not have any ventilator allocated under a limited ventilator supply. When the budget is too tight, the regions with a high transmission rate gets the priority. As the budget increases to medium and ample, the model allocates more capacity to the regions with a higher population and a larger initial number of infections but with a lower transmission rate. Also, the stage-wise distribution of ventilators has a high relationship with the available budget. If the budget is tight, all ventilators are distributed within the first two stages. As we increase the budget, some of the ventilators are allocated in stages three and four in addition to stages one and two. Thus, a higher budget level also provides some flexibility in delaying the ventilator allocation to some regions.

\subsection{Risk Analysis}
In this section, we perform an analysis of the risk parameters $\lambda$ and $\alpha$ in terms of their impact on the expected number of infected and deceased people as well as the CVaR of the impact. Specifically, under the $\$30M$ budget level, we compare four different problems with respect to their risk-averseness level, adjusting $\lambda$ and $\alpha$ values accordingly–risk-neutral  ($\lambda = 0, \alpha= 0$), weak risk-aversion ($\lambda=1, \alpha=0.3$), mild risk-aversion ($\lambda=10, \alpha=0.6$), and strong risk-aversion ($\lambda=10, \alpha=0.95$). The model under the mild risk-aversion results in a high optimality gap ($13\%$) within 7200 CPU seconds running time. Therefore, we solve the scenario-$\omega$ problems described in Section \ref{Cuts} and obtain the lower and upper bounds for the original problem. For our problem, we select five representative scenarios, and add bounds based on the results of those select scenarios in the risk-averse model. After implementing the scenario bounds, the optimality gaps over all of the risk-averseness levels reduce to less than $9.13\%$. 

We decompose the objective function \eqref{objective1-ex:1} into the \textit{Expected Impact} [$\mathbb{E}(f(x, \omega))$] and the \textit{Expected Risk} [$\textrm{CVaR}_\alpha (f(x, \omega))$], as demonstrated in Equation \eqref{eq:mean_risk}, to analyze the impact of risk trade-off on the results. Table \ref{CVaRTable} presents the value of the objective function \eqref{objective1-ex:1}, expected impact, and expected risk (without $\lambda$) under different risk-averseness levels. Specifically, the expected impact represents the expected total number of infections and deceased individuals, and the expected risk corresponds to the expected CVaR term in Eq. \eqref{objective1-ex:1} without the $\lambda$ value. According to Table \ref{CVaRTable}, when both $\lambda$ and $\alpha$ increase, the level of risk-averseness and the expected risk increase. The optimal objective function value increases due to the additional risk  term added into the objective formulation. The expected impact also increases, implying the cost of being risk-averse, which is the increased number of infections and deceased individuals while trying to mitigate specific disastrous scenarios. 

\begin{table}[H] \small
		\renewcommand\arraystretch{0.8}
\renewcommand\tabcolsep{3.0pt}
  \centering
  \caption{Comparison of objective value, expected impact, and expected risk under various risk-averseness levels}
    \begin{tabular}{ccccccccc}
    \hline
		    & \textbf{Risk} & \textbf{Weak} & \textbf{Mild} 			& \textbf{Strong}\\
		    & \textbf{Neutral} & \textbf{Risk-aversion} & \textbf{Risk-aversion} 			& \textbf{Risk-aversion}\\
         &   \textbf{($\lambda=0$, $\alpha=0$)}        &    \textbf{($\lambda=1$, $\alpha=0.3$)}    & \textbf{($\lambda=10$, $\alpha=0.6$)}   & \textbf{($\lambda=10$, $\alpha=0.95$)}\\
    \hline
Objective Value & 347,395 & 721,710 & 3,997,129 & 4,011,964 \\
    \hline
Expected Impact & 347,395 & 360,438 & 362,559 & 363,526 \\
    \hline
Expected Risk  & - & 361,272 & 363,457 & 364,844 \\
    \hline
    \end{tabular}%
  \label{CVaRTable}%
\end{table}%

The expected impact and expected risk (without $\lambda$) for various combinations of $\lambda=\left\{0,1,10\right\}$ and $\alpha=\left\{0.3,0.6,0.95\right\}$ under the $\$30M$ budget level are presented in Table \ref{lambda}. We observe the change of the expected impact and expected risk when changing one of the risk parameters and fixing all others' original values. According to the results, fixing the $\alpha$ value, both expected impact and expected risk show an increasing trend due to the increase of $\lambda$. When we move from risk-neutral ($\lambda=0$) to risk-averse ($\lambda=\left\{1,10\right\}$), the expected impact always increases. Similarly, $\lambda=\left\{1,10\right\}$ increases the expected impact compared to the risk-neutral model. Besides, when fixing the $\lambda$ value and increasing the $\alpha$ value, the expected risk increases because we increase the confidence level for reducing the risk of having an extremely large number of infections and big losses of lives.

\begin{table}[H] \small
		\renewcommand\arraystretch{0.8}
\renewcommand\tabcolsep{3.0pt}
  \centering
  \caption{Expected impact and risk for different risk-averseness levels}

    \begin{tabular}{cccccccccc}
    
    \hline
  \boldmath{$\lambda\backslash\alpha$} & \textbf{\qquad} & \multicolumn{2}{c}{\textbf{0.3}}  & \textbf{\qquad} & \multicolumn{2}{c}{\textbf{0.6}}  & \textbf{\qquad} & \multicolumn{2}{c}{\textbf{0.95}}\\ 
 \cline{3-4}\cline{6-7}\cline{9-10}
 
	& \textbf{\qquad} & \multicolumn{1}{c}{Expected} & \multicolumn{1}{c}{Expected}  & \textbf{\qquad} & \multicolumn{1}{c}{Expected} & \multicolumn{1}{c}{Expected}  & \textbf{\qquad} & \multicolumn{1}{c}{Expected} & \multicolumn{1}{c}{Expected}\\
	\textbf{\qquad}	& \textbf{\qquad} & \multicolumn{1}{c}{Impact} & \multicolumn{1}{c}{Risk}  & \textbf{\qquad} & \multicolumn{1}{c}{Impact} & \multicolumn{1}{c}{Risk}  & \textbf{\qquad} & \multicolumn{1}{c}{Impact} & \multicolumn{1}{c}{Risk}\\
    \hline
    \textbf{0} & \textbf{\qquad} & \multicolumn{1}{c}{347,395} & \multicolumn{1}{c}{0}  & \textbf{\qquad} & \multicolumn{1}{c}{347,395} & \multicolumn{1}{c}{0}  & \textbf{\qquad} & \multicolumn{1}{c}{347,395} & \multicolumn{1}{c}{0}\\         
    
    \textbf{1} & \textbf{\qquad} & \multicolumn{1}{c}{360,438} & \multicolumn{1}{c}{361,272}  & \textbf{\qquad} & \multicolumn{1}{c}{361,950} & \multicolumn{1}{c}{363,251}  & \textbf{\qquad} & \multicolumn{1}{c}{363,882} & \multicolumn{1}{c}{365,236}\\
    
    \textbf{10} & \textbf{\qquad} & \multicolumn{1}{c}{360,880} & \multicolumn{1}{c}{361,341}  & \textbf{\qquad} & \multicolumn{1}{c}{362,559} & \multicolumn{1}{c}{363,457}  & \textbf{\qquad} & \multicolumn{1}{c}{363,526} & \multicolumn{1}{c}{364,844}\\
   
   \hline
    \end{tabular}
  \label{lambda}
\end{table}

\section{Discussion and Future Directions}
In this paper, we present a general multi-stage mean-risk epidemics-ventilator-logistics model and apply this model to control the COVID-19 in select counties of New York and New Jersey. We first explicitly formulate the uncertainty of the proportion of untested asymptomatic infections at each stage, generating a multi-stage scenario tree. We then develop a compartmental disease model and integrate the short-term human movement among multiple regions into this model. We also derive a time- and space-varying disease transmission formulation and a logistics sub-model. We then integrate all those components into one mathematical formulation, which minimizes the number of infections and deceased individuals under different intervention strategies.

We solve the epidemics-ventilator-logistics model under different budget levels to determine the ventilator-distribution optimal timing and location under various pandemic scenarios. Next, we apply the CVaR in a nested form over a five-stage planning horizon to minimize the total expected number of infections and deceased individuals, as well as the weighted risk of the loss. Finally, we solve the scenario sub-problems under various scenarios to generate the lower and upper bounds for the original problem, reducing the optimality gap. Our results provide key insights into the resource-allocation decisions for controlling the COVID-19 and can be adapted to study the transmission and logistics of other similar diseases.

According to the results, the number of infections, deceased individuals, hospitalized individuals, and ICU patients indicates that short-term migration influences the number of infections, even if the transmission rate is constant over time. The impacts of government interventions on the number of deceased individuals, hospitalized individuals, and ICU patients are delayed because deaths and hospitalization have a lag period compared to zero or a small lag phase in the growth of infections. Furthermore, the number of ICU patients at each time period depends on the minimum number of patients who require the ICU and the available ventilators. Thus, the number of ICU patients might be at the capacity limit even under different intervention strategies at a particular stage. The ``Lockdown'' strategy is the best way to control disease transmission. Nevertheless, ``Mask and Social Distance'' applied after the several stages of ``Lockdown'' is the second-best strategy to optimistically alleviate the pandemic's economic impacts.

The ventilator allocation under different budget levels and scenarios indicates that the number of ventilators allocated to each region depends on various factors, such as the number of initial infections, initial disease transmission rates,  initial ICU capacity, and the population of a geographical location. The region with a high initial transmission rate and low initial ICU capacity receives more ventilators under a low disease transmission scenario and a limited budget level. This is because, under a low disease transmission scenario, other regions with low initial transmission rates have fewer infections, even if they have smaller initial ICU capacity. Independent from the budget level, the area with a low initial transmission rate and low initial ICU capacity has more ventilators allocated under the medium transmission scenario. This is because the number of infections in the region with a high initial transmission rate and high initial ICU capacity does not significantly increase even if they receive fewer ventilators under a mild disease transmission scenario. Under a medium and ample budget level, the model allocates more capacity to the regions with a higher population and a larger initial number of infections but with a lower transmission rate. Moreover, when the budget is limited, all of the ventilators are allocated at the first two stages. When the budget becomes ample, decision-makers would have some flexibility in delaying ventilator allocation to later stages of the planning horizon.

The increase in the mean-risk trade-off coefficients in the risk-averse model improves the confidence level, reducing the loss in the right tail of the objective function values (the number of infected and deceased individuals over highly-adverse scenarios). However, we should expect more infections and deceased individuals on average considering all possible scenarios when we want to decrease the impact of adverse scenarios by increasing the risk-averseness level. 

This study leads to several future directions for research. For instance, vaccine allocation is also essential as it can potentially protect people from being infected. The combination of vaccine allocation and other interventions will provide more flexible strategies to prevent and control the disease. For example, for the region with a low transmission rate and high vaccine coverage, decision-makers could consider lifting the ``Lockdown'' earlier to stimulate the economy. Furthermore, the mathematical model cannot allocate ventilators to some regions under a very tight budget, and so future research could investigate ethical and fair resource allocation strategies during a pandemic. Also, some of the assumptions and inferences made in our model could be updated in a future study as more data are available.

\section*{Acknowledgments} 
We gratefully acknowledge the partial support of the National Science Foundation CAREER Award co-funded by the CBET/ENG Environmental Sustainability program and the Division of Mathematical Sciences in MPS/NSF under Grant No. CBET-1554018.

\renewcommand\refname{References} 
\bibliography{MultiSPReferences}

\end{document}